\documentclass[fleqn,usenatbib]{mnras}

\usepackage{caption}
\usepackage{subcaption}
\usepackage{graphicx}
\usepackage{hyperref}
\usepackage{amsmath}
\usepackage[T1]{fontenc}

\DeclareRobustCommand{\VAN}[3]{#2}
\let\VANthebibliography\thebibliography
\def\thebibliography{\DeclareRobustCommand{\VAN}[3]{##3}\VANthebibliography}

\usepackage{graphicx}	% Including figure files
\usepackage{amsmath}	% Advanced maths commands
\usepackage{amssymb}	% Extra maths symbols
\usepackage{cleveref}

\usepackage{newtxtext,newtxmath}

\title[Mixed Stirring of Debris Discs]{A Mixed Stirring Mechanism for Debris Discs with Giant and Dwarf Planetary Perturbations}

\author[Mu\~noz-Guti\'errez et al.]{
Marco A. Mu\~noz-Guti\'errez,$^{1}$\thanks{E-mail: mmunoz.astro@gmail.com (MAM)}
Jonathan P. Marshall,$^{1,2}$
and Antonio Peimbert$^{3}$
\\
$^{1}$Institute of Astronomy and Astrophysics, Academia Sinica, 11F of AS/NTU
Astronomy-Mathematics Building, No.1, Sec. 4, Roosevelt Rd, Taipei 10617, Taiwan\\
$^{2}$Centre for Astrophysics, University of Southern Queensland, Toowoomba, QLD 4350, Australia\\
$^{3}$Instituto de Astronom\'ia, Universidad Nacional Aut\'onoma de M\'exico, Apdo. postal 70-264, Ciudad Universitaria, M\'exico\\
}

\date{Accepted XXX. Received YYY; in original form ZZZ}

\pubyear{2023}

\begin{document}
\label{firstpage}
\pagerange{\pageref{firstpage}--\pageref{lastpage}}
\maketitle

\begin{abstract}

Debris discs consist of belts of bodies ranging in size from dust grains to planetesimals; these belts are visible markers of planetary systems around other stars that can reveal the influence of extrasolar planets through their shape and structure. Two key stirring mechanisms --- self-stirring by planetesimals and secular perturbation by an external giant planet --- have been identified to explain the dynamics of planetesimal belts; their relative importance has been studied independently, but are yet to be considered in combination. In this work we perform a suite of 286 N-body simulations exploring the evolution of debris discs over 1~Gyr, combining the gravitational perturbations of both dwarf planets embedded in the discs, and an interior giant planet. Our systems were somewhat modeled after the architecture of the outer Solar system: a Solar mass star, a single massive giant planet at 30~au ($M_{\rm GP} =$ 10 to 316~$\mathrm{M}_{\oplus}$), and a debris disc formed by 100 massive dwarf planets and 1\,000 massless particles ($M_{\rm DD} =$ 3.16 to 31.6~$\mathrm{M}_{\oplus}$). We present the evolution of both the disc and the giant planet after 1~Gyr. The time evolution of the average eccentricity and inclination of the disc is strongly dependent on the giant planet mass as well as on the remaining disc mass. We also found that efficient stirring is achieved even with small disc masses. In general, we find that a mixed mechanism is more efficient in the stirring of cold debris discs than either mechanism acting in isolation.

\end{abstract}


%%%%%%%%%%%%%%%%%%%%%%%%%%%%%%%%%%%%%%%%%%%%%%%%%%

%%%%%%%%%%%%%%%%% BODY OF PAPER %%%%%%%%%%%%%%%%%%

\section{Introduction}

Debris discs are massive structures observed around 20 to 30 percent of main sequence stars \citep[for recent reviews see, e.g.,][]{Wyatt18,Hughes18}; their presence is signaled by the presence of excess emission in thermal emission at infrared to millimetre wavelengths \citep[e.g.,][]{Eiroa13,Thureau14,Holland2017,Sibthorpe18} and/or scattered light at optical or near-infrared wavelengths \citep[either total intensity or polarization, e.g.,][]{Schneider14,Esposito20}, coming from circumstellar micrometre- to centimetre-sized dust grains. 

The dust contents of debris discs are not just remnants of the original, massive, dust- and gas-rich protoplanetary discs of material from which planets are born \citep{Wyatt15,2018Andrews}. Although some amount of (sub-)micron-sized dust can remain after the initial protoplanetary disc dissipates, the smallest dust grains are lost on timescales much shorter than the age of the host star due to photoevaporation and accretion processes \citep{BurnsLamySoter79,Krivov10}. Therefore, the dust observed in debris discs is thought to be second-generation dust, produced in disruptive collisions between larger leftover planetesimals, which were originally formed from dust (and ices) in the protoplanetary discs. Collisions between these bodies produce detectable amounts of dust throughout the lifetime of the host star and beyond \citep[e.g.,][]{Matthews14,Farihi16}. However, to be able to produce that dust, planetesimals must be abundant enough to have frequent collisions, as well as have relative velocities high enough for collisions to be destructive, or at least erosive \citep{Dohnanyi69,KenyonBromley01}.

The formation of planetesimals starts with dust growth in protoplanetary discs, which is encouraged by vertical settling of larger grains to the disc mid-plane and radial trapping at pressure bumps, especially around ice lines increasing the mass surface density to a level where the gas-to-dust ratio approaches unity \citep{2008Blum,2017Drazkowska}. Growth beyond millimetre- to centimetre-sized particles is inhibited by collisions due to the `bouncing barrier' \citep{2008Brauer,2010Birnstiel,2010Zsom}. The rapid loss of these large grains or pebbles due to inward radial drift is an inhibiting factor in current theories of planet formation. A mechanism referred to as the `streaming instability' has been proposed as a means to bypass the `bouncing barrier' and precipitate planetesimals directly from pebbles in the proto-planetary disc \citep{2005Youdin,2007Youdin,2007Johansen,2010aBaiStone,2010bBaiStone}. The size distribution of these bodies is consistent with the range observed in the Solar System's Kuiper Belt, wherein Pluto and its cohort could represent the high mass tail of this planetesimal formation process \citep{Johansen15,2016Simon}.  

The initial orbits of planetesimals formed in the protoplanetary phase are expected to be nearly circular and confined to the disc mid-plane, therefore some additional stirring mechanism is required to dynamically excite the planetesimal belts left after the gas dispersal. Structures observed in proto-planetary discs, such as rings, spiral arms, etc., are uncorrelated with ice lines/density enhancements induced by disc temperature structure \citep{2018Long,2019vanderMarel}. Rings in protoplanetary discs could therefore be the result of the action of protoplanets trapping material and sculpting the disc \citep[e.g.][]{Dong18,Huang18,Zhang18}.Low mass companions have been identified embedded within several such discs \citep{2018Fedele,2019Keppler,2020Ubeira-Gabellini,2021Teague}. Once the eccentricity-damping effect of the protoplanetary gas disc has been removed, the ongoing stirring by either planets or planetesimals on the debris disc will excite the belt leading to enhanced collision rates. 

Inheritance of structure from proto-planetary discs to debris discs is uncertain \citep{2022Najita}, but planetesimal belt locations in cold debris discs (exoKuiper belts) appear consistent with formation at CO ice line \citep{2018Matra,2021Marshall}. However, the widths of rings in proto-planetary discs are much narrower than debris disc's planetesimal belts \citep{2021Miller}. The majority of broad debris belts observed by ALMA with sufficient spatial resolution exhibit sub-structures consistent with the presence of a perturbing planetary companion \citep{2021Marino}.

Analysis of spatially resolved observations of debris discs have been used to infer the stirring mechanism(s) in play for a number of young systems based on stirring arguments from the size of the disc and the stellar age \citep[e.g.][]{Moor15,Vican16} and interpretation of their architectures, revealing disc-planet interactions in a variety of ways, including the detection of gaps in broad belts \citep[e.g.][]{Marino17,Marino18,Macgregor19}, scattered haloes of mm dust grains \citep{2018Macgregor,Geiler19}, and the eccentric architectures of narrow belts \citep{2020Kennedy}. Most recently, \cite{Pearce22} examined a large ensemble of debris discs, both spatially resolved and unresolved, inferring the required mass of a perturber, under the assumption that the sculpting is produced by a single planet or multiple planets, as well as if being the result of self-stirring by massive planetesimals within the disc.

These two aforementioned main mechanisms have been suggested in the past to account for the planetesimal excitation levels, i.e. 1) the self-stirring mechanism \citep[e.g.][]{Kenyon08,Krivov18}, in which large planetesimals are able to trigger a collisional cascade once they acquire a certain size threshold, and 2) the secular perturbations from giant planetary companions, interior or exterior to the discs \citep[e.g.,][]{Wyatt99,Mustill09}. The latter has been favored recently due to the very large masses of debris discs required to explain their excitation levels by the self-stirring mechanism \citep{KrivovWyatt21,Pearce22}.

However, the effects of a simultaneous stirring by external planets together with internal planetesimals has never been studied in detail. Besides, the existing self-stirring models do not properly account for the top-end of the size distribution (e.g., Pluto-sized dwarf planets), frequently relying in models comprised of equal massed (not so large) bodies stirring the disc. 

In previous works \citep{Munoz15b,Munoz17b,Munoz18}, we studied the long-term evolution of generic cold debris discs of different masses, under the perturbations of an interior Neptune-like giant planet, as well as of dozens of dwarf planet-sized massive perturbers (DPs, hereafter) embedded in the discs. In \citet{Munoz17b}, we demonstrated the existence of a stabilizing effect produced by a giant planet over the disruptive perturbations of massive DPs; we also demonstrated \citep{Munoz18} the existence of a constant resupplying of the giant's MMRs with new objects, a mechanism acting on secular time-scales due to the radial migration of disc particles produced by the DPs' scattering effects.

In this work, we expand the exploration of the mass parameter space of our mixed stirring scenario for more massive debris discs, comparable to those which have been observed in extrasolar planetary systems. We account for both the perturbations produced by an interior giant planet, as well as 100 massive DPs embedded in a disc of 1\,000 massless particles. The simulation setup for our grid of disc-planet systems, along with a brief summary of the dynamical modeling approach, is given in \Cref{sec:method}. In \Cref{sec:results} we characterize the outcome of our simulations, through analysis of the evolution of the survival fraction, average eccentricities, and inclinations of the bodies comprising the discs, as well as the orbital perturbations exerted on the giant planet; we provide our interpretation of the results and how they relate to other works addressing either planetary or self-stirring of a debris disc in isolation in \Cref{sec:discussion}. Finally, in \Cref{sec:conclusions}, we summarize our findings and present our conclusions.

\section{Methods and Simulations}
\label{sec:method}

We aim to test the efficiency for producing stirring over debris disc particles, of models which combine the perturbations coming from a giant planet, located interior to an initially wide and cold debris disc, as well as dwarf planets embedded within the disc. We call this a mixed stirring scenario, since it combines some of the elements applied so far in debris discs stirring models, i.e. secular perturbations from giant planets and self-stirring. 
%initial grid used 750, 2nd grid used 700, 3rd grid 450

\subsection{Model disc generation}

Our systems are formed by a Solar mass central star, as well as a Neptune-analog ``giant'' planet (GP, hereafter) located at 30~au and starting with zero eccentricity and inclination. The debris disc is formed by 1\,000 test particles and 100 massive DPs; the disc is 30~au wide and its inner edge is set to be 10 Hill radii beyond the GP location. We assume the mass of the debris disc to be given by the sum of the individual masses of the 100 DPs.

We study a grid of models where the GP mass explores values from 10 to $316~\mathrm{M}_{\oplus}$ (i.e. from sub-Neptune to one Jupiter masses), in logarithmic steps of 0.15 (11 values). 

The mass of the debris discs covers a range from 3.16 to 31.6~$\mathrm{M}_{\oplus}$ in logarithmic steps of 0.04 (26 values). 

Within each disc, the masses of the individual DPs are drawn randomly to try to reproduce the 100 most massive particles of a mass distribution $n(m) \propto m^{-2.8}$, consistent with the distribution of large bodies in the Kuiper Belt \citep{Fraser09}. The individual mass of the most massive DP in the lightest disc is below 0.105 $\mathrm{M}_{\oplus}$, while in the most massive disc it is 1.05 $\mathrm{M}_{\oplus}$. Those values correspond to ratios with the less massive giant planet of 0.01 and 0.1, respectively. Such large planetesimal masses are not unexpected according to recent theories of planetesimal formation \citep[e.g. the streaming instability;][]{2005Youdin,Morbidelli09,Nesvorny19}, and are consistent with recent measurements of planetesimal masses inferred from the spatially resolved scale heights of $\beta$ Pic and AU Mic \citep{2019Matra,2019Daley}.

The range in debris disc masses was chosen to keep a realistic representation of the individual objects in the discs while remaining computationally feasible, i.e. a larger range in debris discs masses would imply that individual DPs would be very massive (comparable to the GP mass) to account for more massive discs, or we would require to proportionally increment the number of DPs in our simulations, making them too computationally expensive. If lower limits on the DP masses are preferred, the largest of these DPs should be interpreted as the sum of many smaller bodies, a product of the limitation of our computational power.

The distributions of semi-major axes, eccentricities, and inclinations of the DPs and test particles were randomly generated based on a single seed. The initial inclinations of the DPs and test particles were randomly drawn between 0 and $5\degr$, whilst the initial eccentricities were constrained to be $\leq$0.05, i.e. we used similar values to the ones found for the cold classical Kuiper Belt \citep{Gulbis10}. Visual inspection of the output for 20 seed values was carried out and an initial simulation setup was selected based on the uniformity of the distribution in $a$-$e$ and $a$-$i$ parameter space for both the DPs and test particles \footnote{The initial orbital distribution of the DPs and test particles in the discs, for the random seed used in this work, can be found online at \href{https://figshare.com/projects/Mixed_Stirring_of_Debris_Discs/136118}{Figshare}.}.

\subsection{N-body simulations}

We used the hybrid symplectic integrator from the {\sc mercury} package \citep{Chambers99}, to explore the long-term evolution of a grid of 286 debris disc models. An initial time-step of 400 days is used in all cases, as well as an accuracy parameter for the Bulirsch-Stoer integrator of $10^{-10}$. We produced orbital outputs every 10~Myr, over a total integration time of 1~Gyr.

Particles are removed from the simulation if their semi-major axes grow larger than 10\,000~au, decrease below 1~au, or if they collide with the GP or the DPs. In most cases, several DPs are also ejected from the simulations by the same mechanisms due to their mutual interactions.

\section{Results}
\label{sec:results}

Over sufficiently long periods of time ($\sim$100~Myr), the gravitational perturbations from DP-sized objects, acting on initially cold debris discs particles, induce a considerable vertical and radial heating \citep{Munoz15b}, which results in a progressive increment of the disc's mean eccentricities and inclinations. 

A GP in a non-circular, non-planar orbit, will induce secular perturbations on an external debris disc, forcing a component on the particles' eccentricity and inclination vectors \citep[e.g.][]{MurrayDermott99,Mustill09,GladmanVolk21}. Though initially circular and planar, the orbit of the GPs in our simulations quickly evolves, as we will show, due to their interactions with the massive disc members (DPs), which makes the former phenomenon relevant. Moreover, under the right circumstances, i.e. if massive enough ($\gtrsim$100~$\mathrm{M}_{\oplus}$), an interior GP can also act to stabilize the orbits of massless particles within debris discs, acting against the perturbations produced by DPs \citep{Munoz17b}.

In the following subsections, we will show separately the evolution of the populations of massless particles, DPs, the debris discs as a whole, and finally the GPs within these systems.

\subsection{Evolution of Massless Particles in the Discs}
\label{ssec:tpsevol}

\begin{figure}
    \centering
        \begin{subfigure}{0.48\textwidth}
        \centering
        \includegraphics[width=\linewidth]{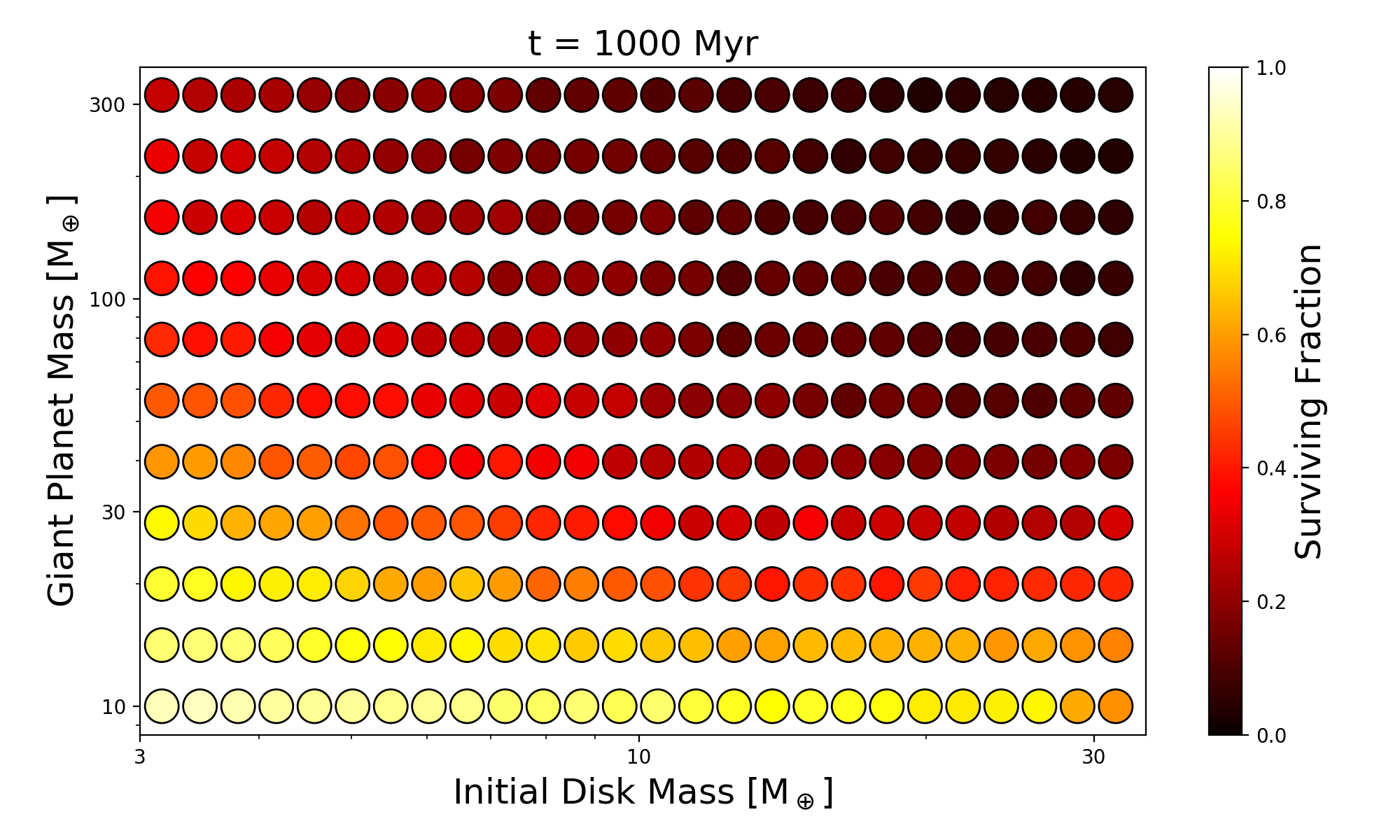}
        \end{subfigure}
    \begin{subfigure}{0.48\textwidth}
        \centering
        \includegraphics[width=\linewidth]{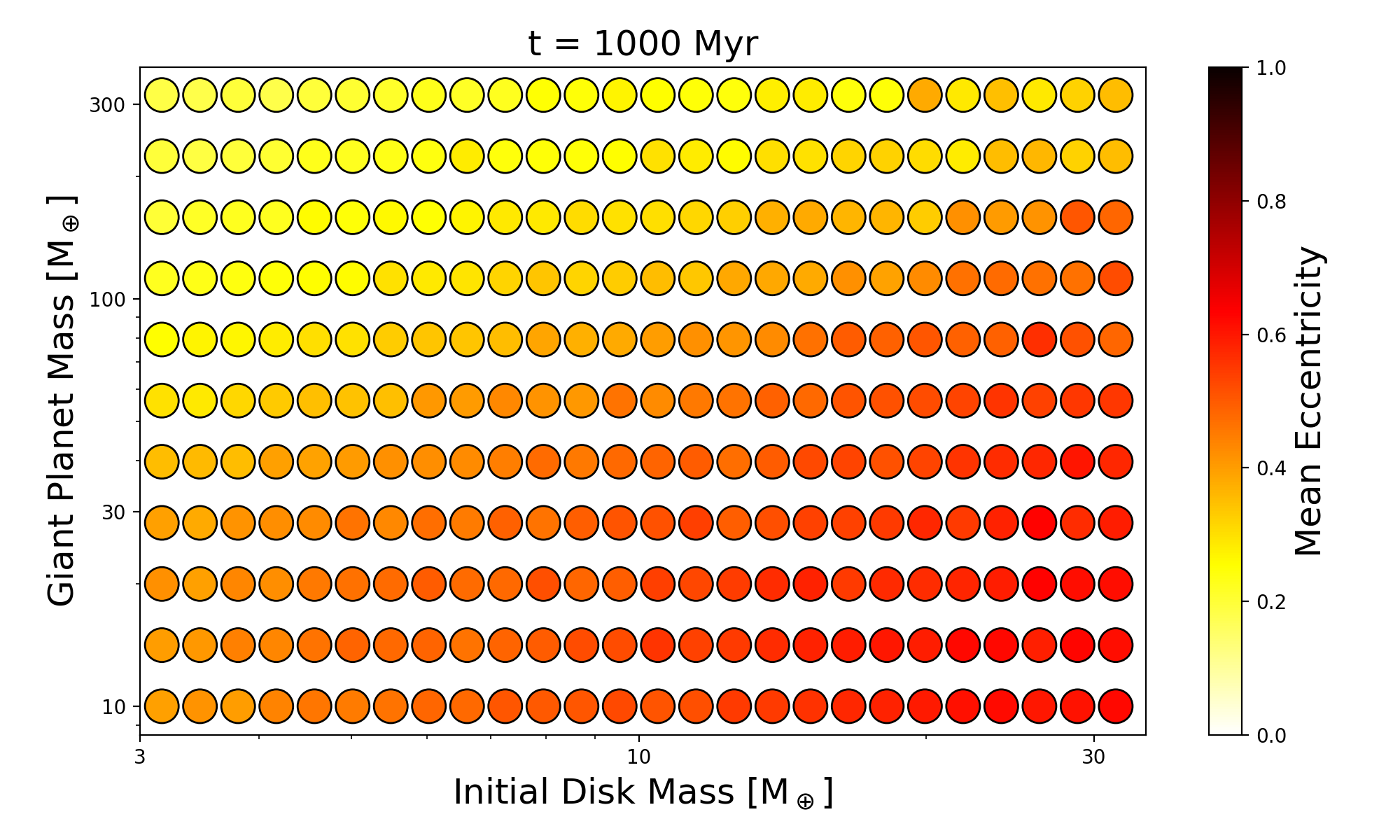}
    \end{subfigure}
    \begin{subfigure}{0.48\textwidth}
        \centering
        \includegraphics[width=\linewidth]{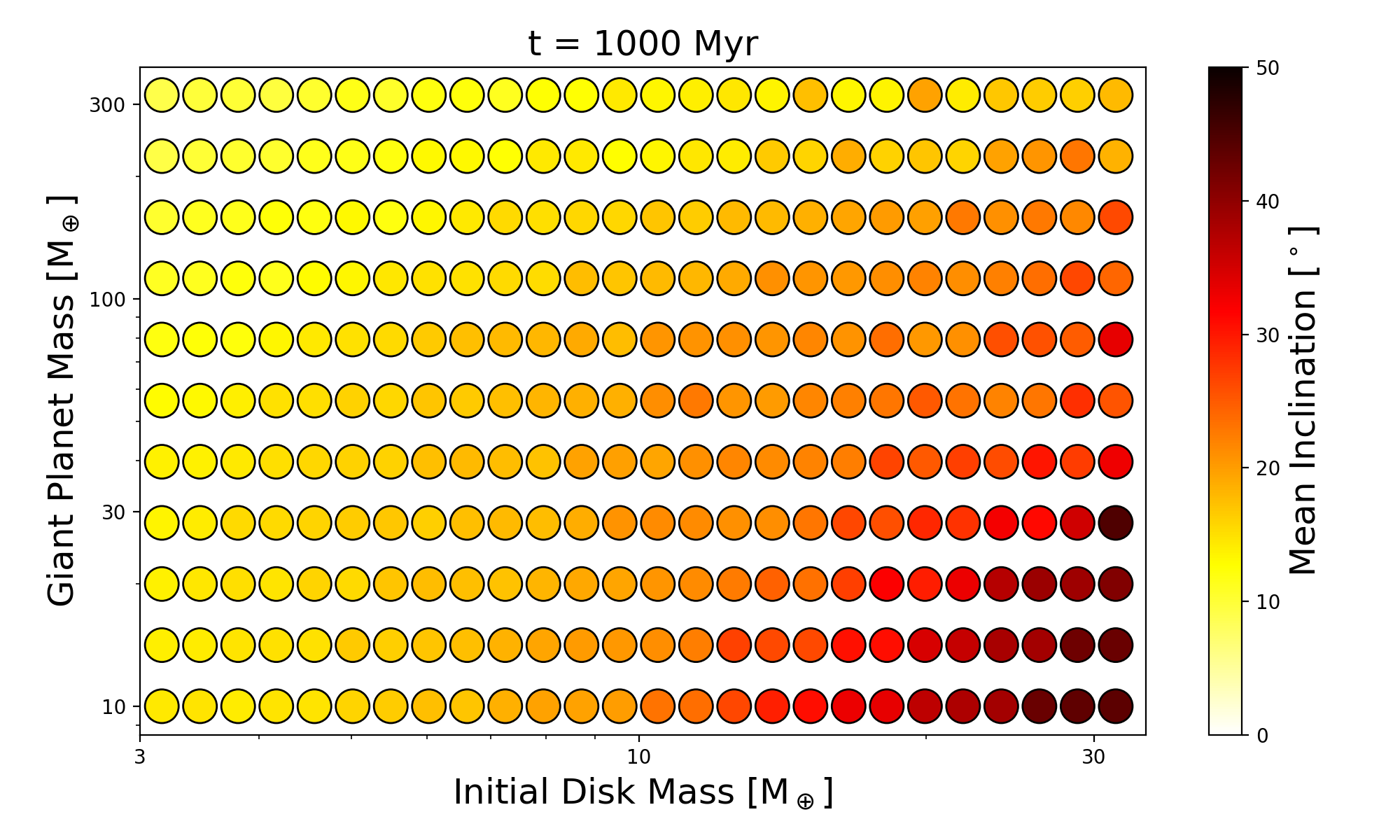}
    \end{subfigure}
    \caption{Animated figures for the survival fraction of test particles in the discs (top), as well as their mean eccentricity (middle) and inclination (bottom). Each colored circle in the grids shows the corresponding value at each time step output from the simulations (i.e. every 10~Myr) according to the color bar presented to the right of each grid. The points in the grid are arranged as a function of the mass of the GP in the model as well as of the initial mass of the debris disc, as accounted by the total mass of 100 massive DPs. The still frames in each panel show the final states of the simulations after 1 Gyr (An animated version of this figure can be found online at \href{https://figshare.com/projects/Mixed_Stirring_of_Debris_Discs/136118}{Figshare}.).}
    \label{fig:gridIniMass}
\end{figure}

We aim to quantify the long-term impact that the combination of perturbers, namely an interior GP plus 100 massive embedded DPs, have on the dynamical stirring of the initially cold debris disc particles. We produced coloured grids showing the survival fraction of particles in the discs, as well as the amount of dynamical excitation, characterised by their mean eccentricities and mean inclinations; at this first stage, we characterise this excitation as a function of the total \textit{initial} disc mass, as well as the GP mass.

In \cref{fig:gridIniMass} we show the evolution of the survival fraction, the average eccentricities, and the average inclinations of the massless particles on our array of simulations.

In the top panel of \cref{fig:gridIniMass} we show the evolution of the survival fraction up to 1~Gyr. In the animated figure, each snapshot corresponds to a 10~Myr time step. The color of each circle represents the surviving fraction of massless particles within that disc, while its location on the grid corresponds to the initial mass of the disc (i.e. the sum of the masses of our DPs) and the mass of the GP in that planetary system.

The ejection efficiency is correlated to both the GP mass and the (initial) disc mass. Those systems with the highest GP and disc masses are the most quickly depleted. Within the first 30 to 100~Myr, the systems with the most massive GPs ($M_{\rm GP} > 100\mathrm{M}_{\oplus}$) have already lost $\geq 80\%$ of their initial particles. Over the next hundreds of Myr, with a smaller number of total particles as well as a smaller number of total perturbers, the ejection rate slows down. Overall the most efficient ejection continues to occur in systems with simultaneously the most massive discs and the most massive GPs.

At the end of the simulations, the higher ejection efficiency occurs for initial discs masses $\gtrsim$10~$\mathrm{M}_{\oplus}$, with the highest ejection efficiency occurring when the GP mass is $\sim$ 100~$\mathrm{M}_{\oplus}$ and the disc mass is $\gtrsim$ 20~$\mathrm{M}_{\oplus}$. Many systems exhibit the ejection of a substantial fraction of the test particles in our simulations. The average ejection rate is 64.5\% across all simulations in the grid, with an ejection rate of up to 96.3\% for the most extreme case. 

The orbital characteristics (eccentricity, inclination) of particles in the discs were calculated by averaging the elements of surviving particles at each time step. In the animated version of \Cref{fig:gridIniMass}, we show their evolution in 10~Myr time steps illustrating the change in the remaining particles, their eccentricity, and inclination over 1~Gyr. The color of each circle represents the mean values of the eccentricity and inclination for each model at the timestep in question.

From the evolution seen in the middle and bottom panel animations of \cref{fig:gridIniMass}, we find that the disc response is monotonic for the lower GP masses ($M_{\rm GP} \leq 30~M_{\oplus}$). We find increasing excitation for decreasing GP mass, increasing disc mass, and longer integration times. The evolution of the disc excitation with time is more clearly visible in eccentricity than inclination.

The middle panel of animated \Cref{fig:gridIniMass} shows that after a few tens of Myr of evolution, a more efficient stirring has been produced for the middle rows of the grid, i.e. for $M_{\rm GP}$ in the range $\sim$30 to $\sim$110~$\mathrm{M}_{\oplus}$. At this time, the stirring grows in proportion to the debris disc mass, while for a given debris disc mass, the stirring increases with GP mass, reaches a maximum around 70 to 100~$\mathrm{M}_{\oplus}$, and decreases for larger GP masses. This behaviour does not resemble the quadratic behaviour presented in \citet[][]{Munoz17b}, however that study was for discs 2 to 4 orders of magnitude lighter than what we are studying here.

Over time, during the first 400 Myr, we see less massive GPs becoming progressively more efficient at exciting test particles; while the more massive GPs models stop evolving. After 300 Myr the sweet spot for efficient stirring becomes less evident, in part due to the ejection rate of the most excited particles from these systems; after 600 Myr even the models with the least massive GP have stopped evolving. By the end of the simulations, the largest mean eccentricity occurs in the lower right corner of the grid, where the disc masses are comparable to, or even greater than, the GPs masses in these systems. 

In the bottom panel of animated \cref{fig:gridIniMass} we observe a slower and more linear trend for the evolution of the mean inclination; up to 100~Myr, the increment in mean inclination is small and its value remains almost homogeneous across the grid. With time a small tendency of larger excitation with larger disc masses and smaller GP masses starts to develop; after 200 Myr the lower right corner of the grid, where $M_{\rm GP} \leq M_{\rm DD}$, starts to show clear signs of a stronger stirring. By the end of the simulations, the final stirring is shown to be a function of both GP mass and debris disc mass, with the greater stirring observed in systems with lower GP masses and larger disc masses. 

When the mass of the disc is comparable to that of the GP, the planet-disc interactions are warranted to be complex. The angular momentum that can be transferred from the GP to the DPs is large enough to produce a significant migration of the GP due to the ejection of massive objects. Also, the reference plane (or ``invariable plane'') within such a massive debris disc is not well defined, as the GP orbit no longer plays such an important role in determining the total angular momentum of the system. These conditions are satisfied for models in the lower right corner of our grid; in that region, particles are excited but they are not efficiently ejected, so the system effectively heats up and there is no way of cooling it down.

\begin{figure}
    \centering
        \begin{subfigure}[t]{0.445\textwidth}
        \centering
        \includegraphics[width=\linewidth]{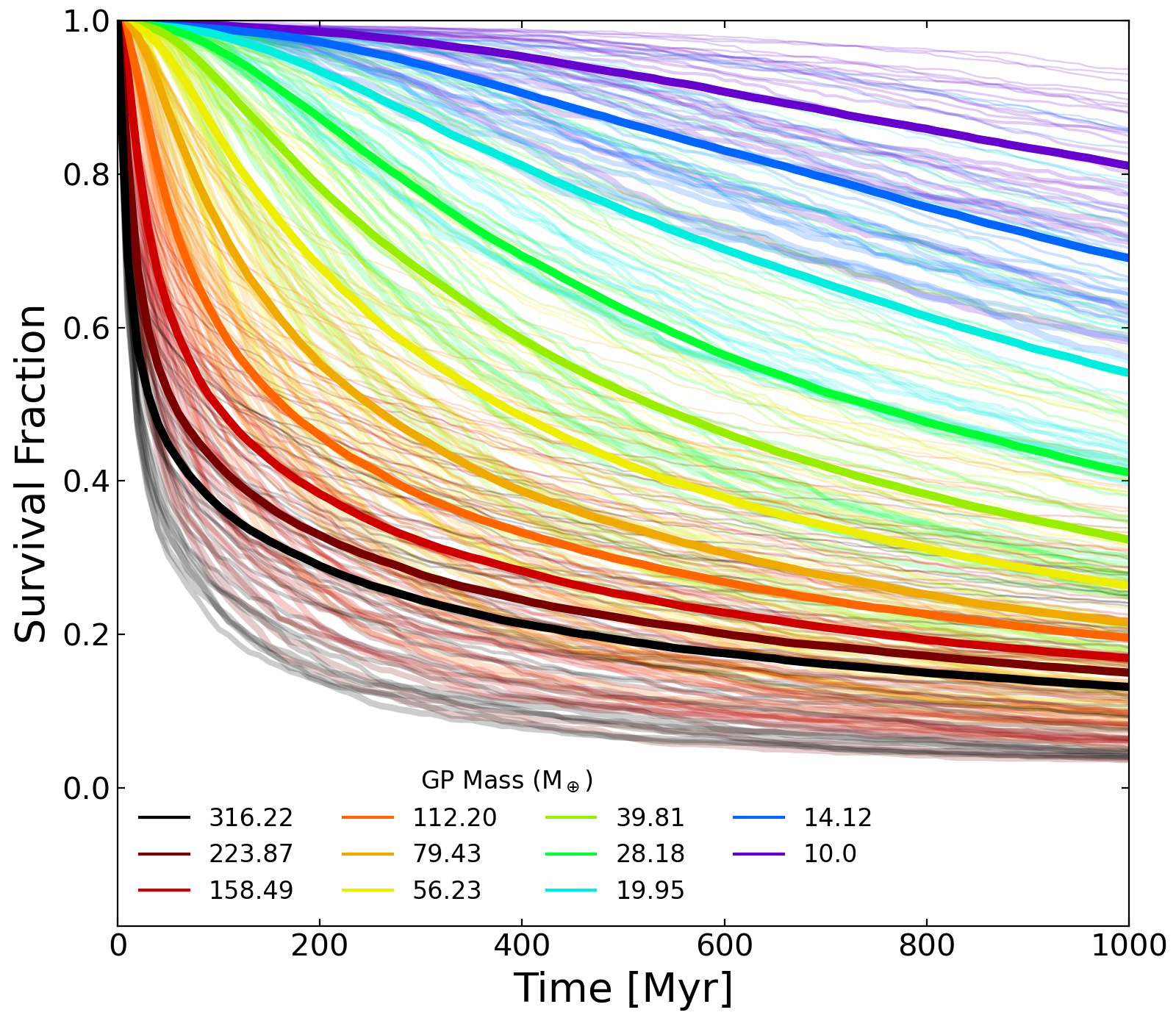}
        \end{subfigure}
    \begin{subfigure}[t]{0.445\textwidth}
        \centering
        \includegraphics[width=\linewidth]{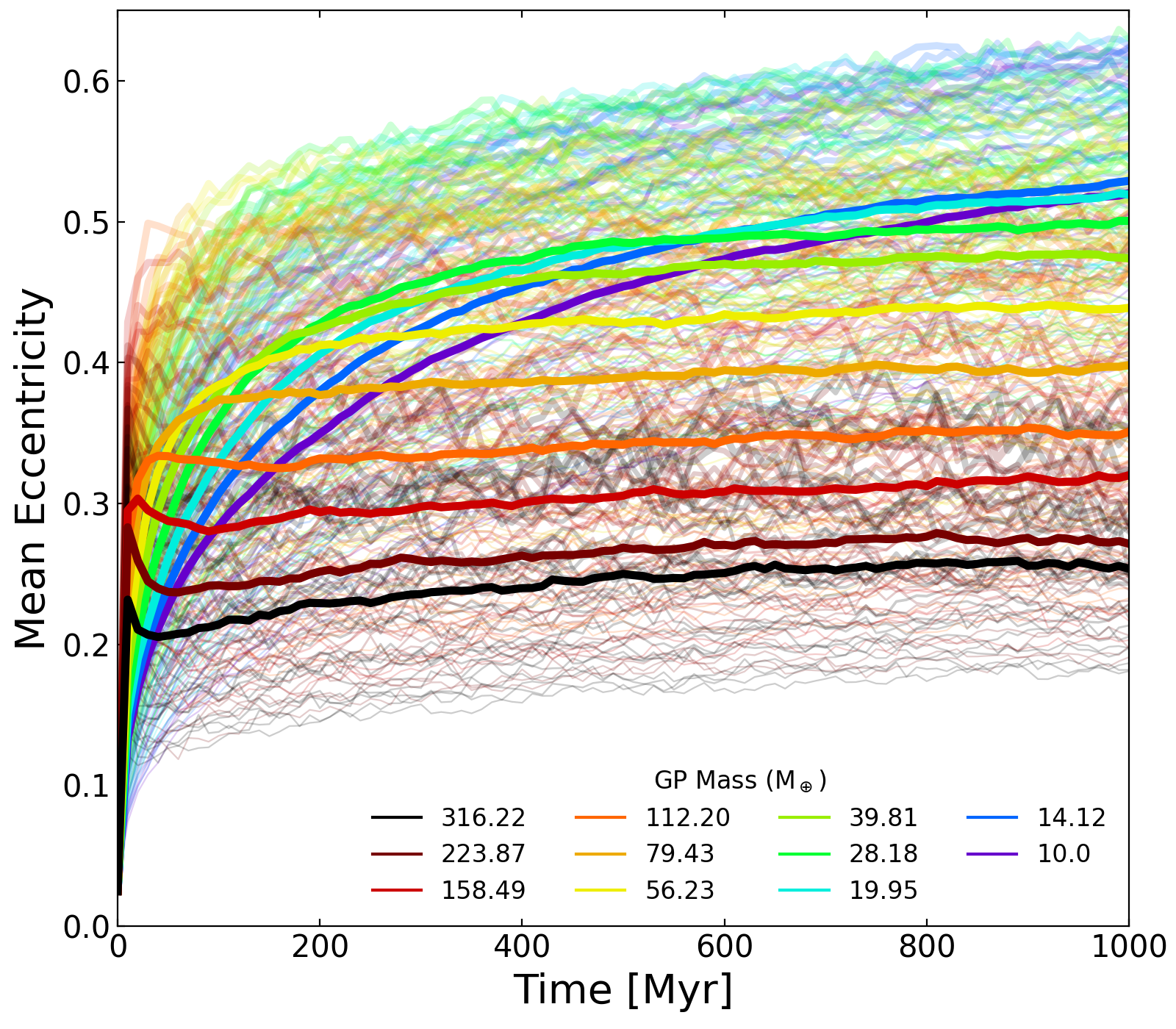}
    \end{subfigure}
    \begin{subfigure}[t]{0.445\textwidth}
        \centering
        \includegraphics[width=\linewidth]{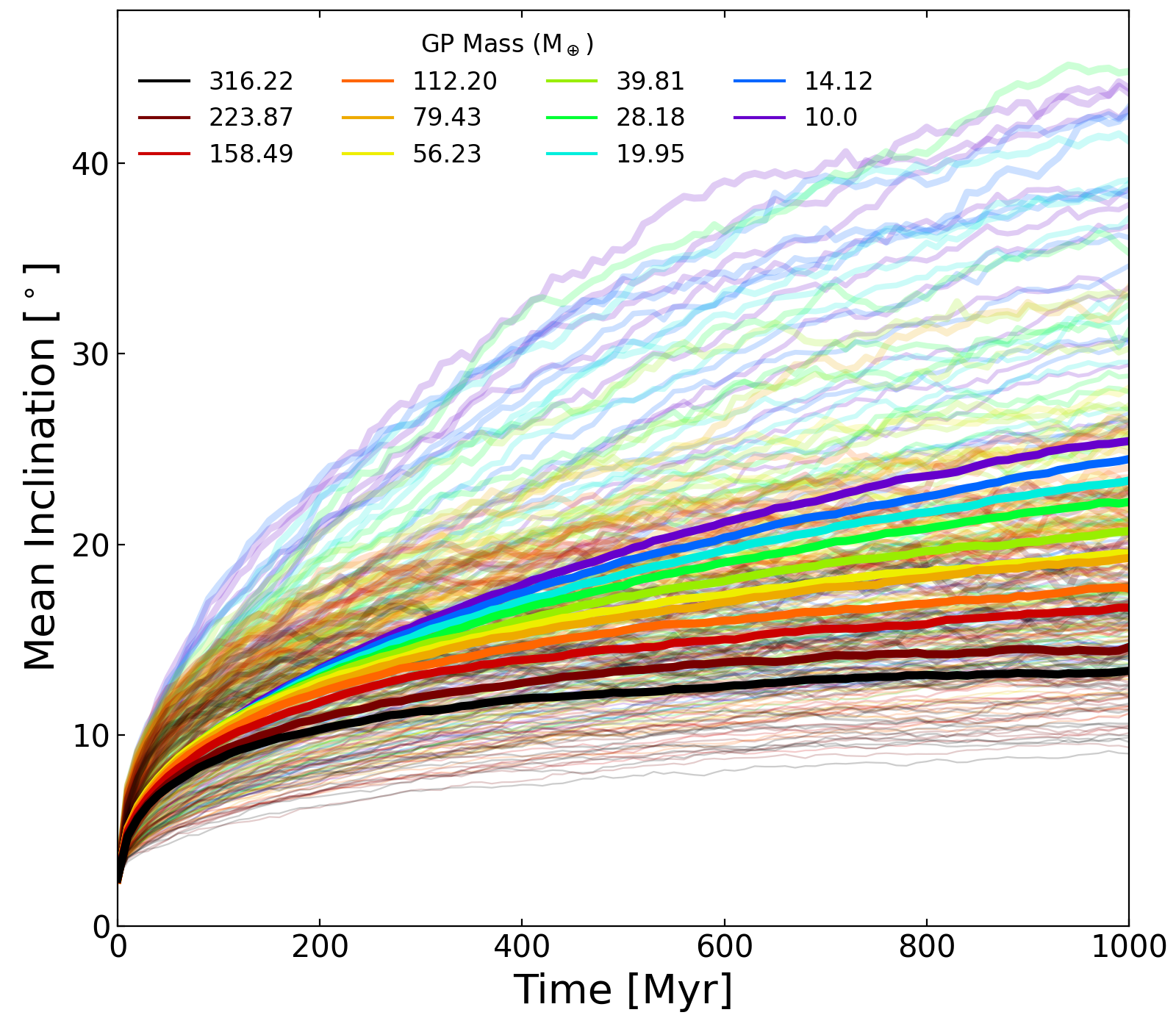}
    \end{subfigure}
    \caption{Evolution of the survival fraction (top), mean eccentricity (middle), and inclination (bottom) of test particles in the discs. The different colors of the lines in the three panels indicate the mass of the GP in the models. The initial debris disc mass in the models is represented by the thickness of each line, with thicker lines corresponding to more massive discs (pale lines in the still frames, all but the last frame in the animated figures). The thickest lines in the still frames (and those of the last animated frames) correspond to the average of all disc masses for any given GP mass (An animated version of this figure can be found online at \href{https://figshare.com/projects/Mixed_Stirring_of_Debris_Discs/136118}{Figshare}).}
    \label{fig:evol}
\end{figure}

Complementary to the animated grids in \cref{fig:gridIniMass}, we also present the time evolution of each model as a curve on the three panels of animated \cref{fig:evol}. There we can see the evolution of each model across the animation, with the survival fraction on the top panel, mean eccentricity in the middle panel, and mean inclination in the bottom panel; the last images (as well as the still frames) highlight the average of all the models with the same GP mass. 

In the top panel of animated \cref{fig:evol}, we see the decline in particle numbers as a function of time. As expected, the more massive planets are more efficient at ejecting test particles from the system. For a given planet mass, the ejection is more efficient with a more massive disc. In the middle panel of animated \cref{fig:evol} we present the eccentricity evolution for each of our 286 models; as in \cref{fig:gridIniMass}, we are presenting the evolution of the mean eccentricity of all particles remaining in the simulations. For models with GP masses less than $\lesssim80~\mathrm{M}_\oplus$ we can see that the eccentricities keep increasing over the whole duration of most of the simulations; all models slow down with time, but for models with GP masses between $\sim30~\mathrm{M}_\oplus$ and $\sim80~\mathrm{M}_\oplus$ there seem to be two phases: first, a fast increase and then they reach a plateau with very little increase in eccentricity thereafter; the change between these two phases occurs sooner and at a lower average eccentricity for the more massive GPs, and will likely occur even at GP masses less than $30~\mathrm{M}_\oplus$, but it probably requires more than 1~Gyr for the same to happen, while for $80~\mathrm{M}_\oplus$ it only requires approximately 100~Myr. For the most massive GPs ($\gtrsim100~\mathrm{M}_\oplus$) a third phase appears, after the fast increase, and before the plateau, a moderately fast decrease occurs due to the rapid ejection of the most eccentric objects; again the evolution is faster for more massive GPs, this new phase seems to be most pronounced for our $223~\mathrm{M}_\oplus$ models, but perhaps with smaller time steps it might be even more important for the $316~\mathrm{M}_\oplus$ GP. Finally, models with GPs more massive than $220~\mathrm{M}_\oplus$ seem to reach saturation, perhaps even a small decline, near the end of the simulations.

Regarding the effect of the disc mass on the overall eccentricity, we find that, for a given time and GP mass, larger disc masses produce larger mean eccentricities.

We present the inclination evolution in the bottom panel of animated \cref{fig:evol}; as for eccentricity, we are presenting the evolution of the mean inclination of all particles remaining in the simulations. Here we show that the evolution of the inclinations is much slower than for the eccentricities, in fact, the inclination for all models continues to rise until the end of the simulations. As with eccentricities, simulations with more massive discs tend to evolve faster and have larger mean inclinations.

In general, for very large GP masses, both eccentricity and inclination show a mostly smooth evolution. This is related to the dominance of the GP mass on the overall dynamics, as well as to the number of particles quickly ejected from the system. This shows a dependence in GP mass on the degree of stirring of the disc. Very massive GPs become less efficient with time at heating the discs, and in fact, those discs cool off at later times, whereas less massive GPs continually stir their discs throughout the timescale of the simulations. This effect can be explained through the ejection efficiency of the GPs at different masses. High-mass GPs (top rows) quickly excite and eject disc particles and DPs that stray into regions of strong interaction with the GP, leaving a depleted but dynamically cold system in their wake. In contrast, low-mass GPs do very little to stir the discs, but also very little to suppress stirring by the DPs or to eject particles excited by DPs, leaving a well-populated but dynamically hot system.

\subsection{Evolution of DPs in the Discs}
\label{ssec:DPsevol}

\begin{figure}
    \centering
    \begin{subfigure}[t]{0.445\textwidth}
        \centering
        \includegraphics[width=\linewidth]{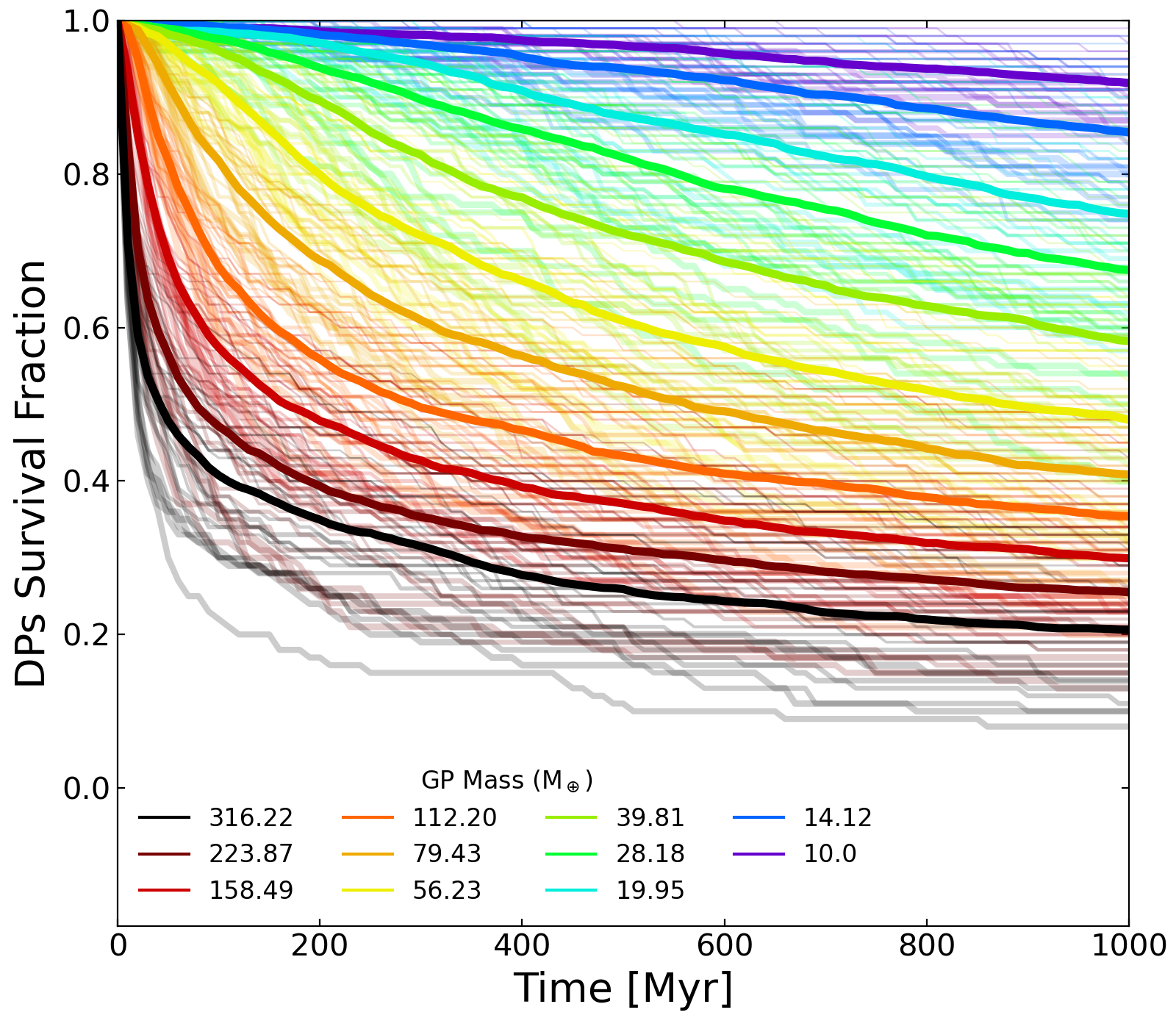} 
%        \caption{Eccentricity} 
    \end{subfigure}
    \begin{subfigure}[t]{0.445\textwidth}
        \centering
        \includegraphics[width=\linewidth]{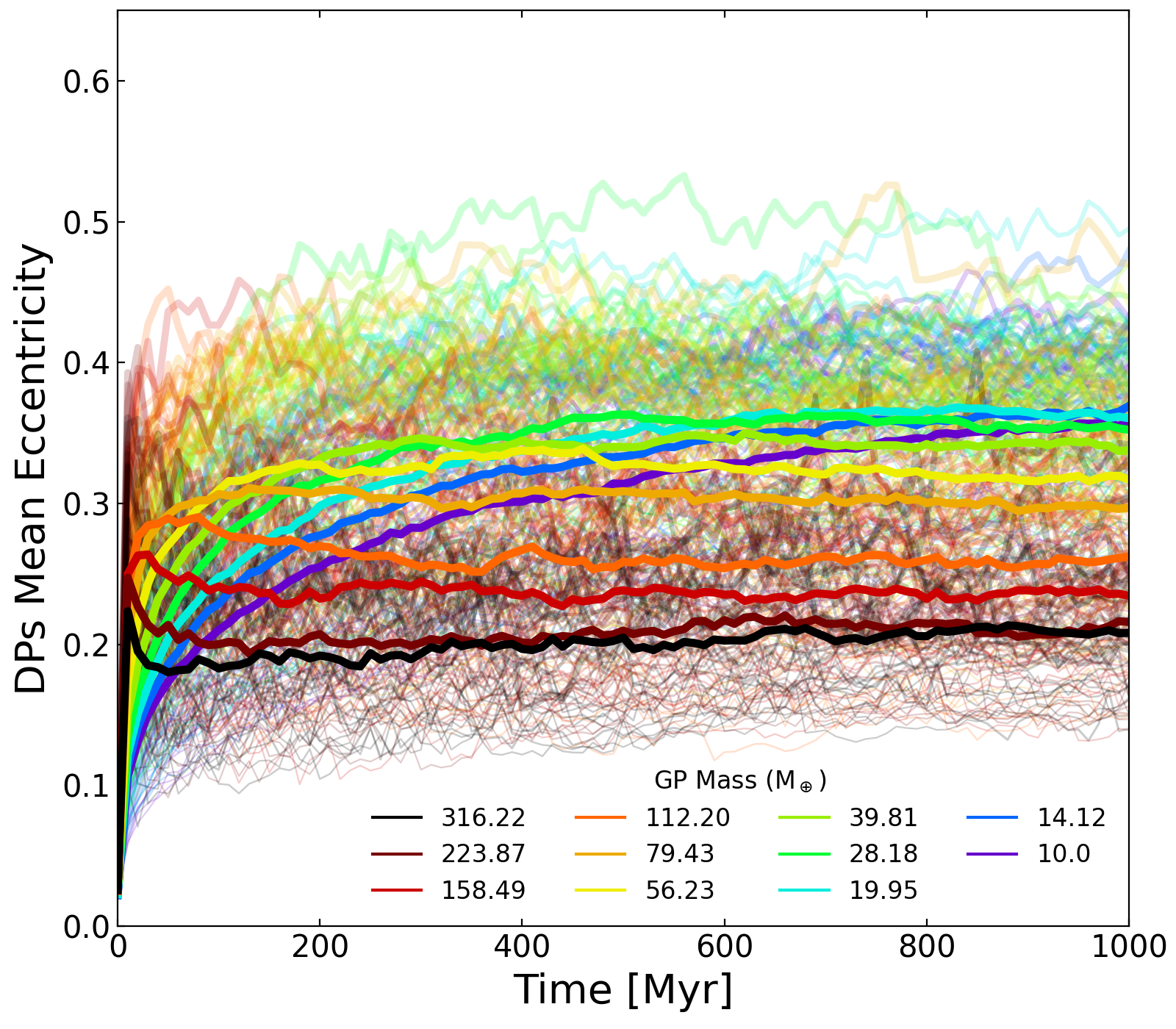} 
%        \caption{Eccentricity} 
    \end{subfigure}
    \begin{subfigure}[t]{0.445\textwidth}
        \centering
        \includegraphics[width=\linewidth]{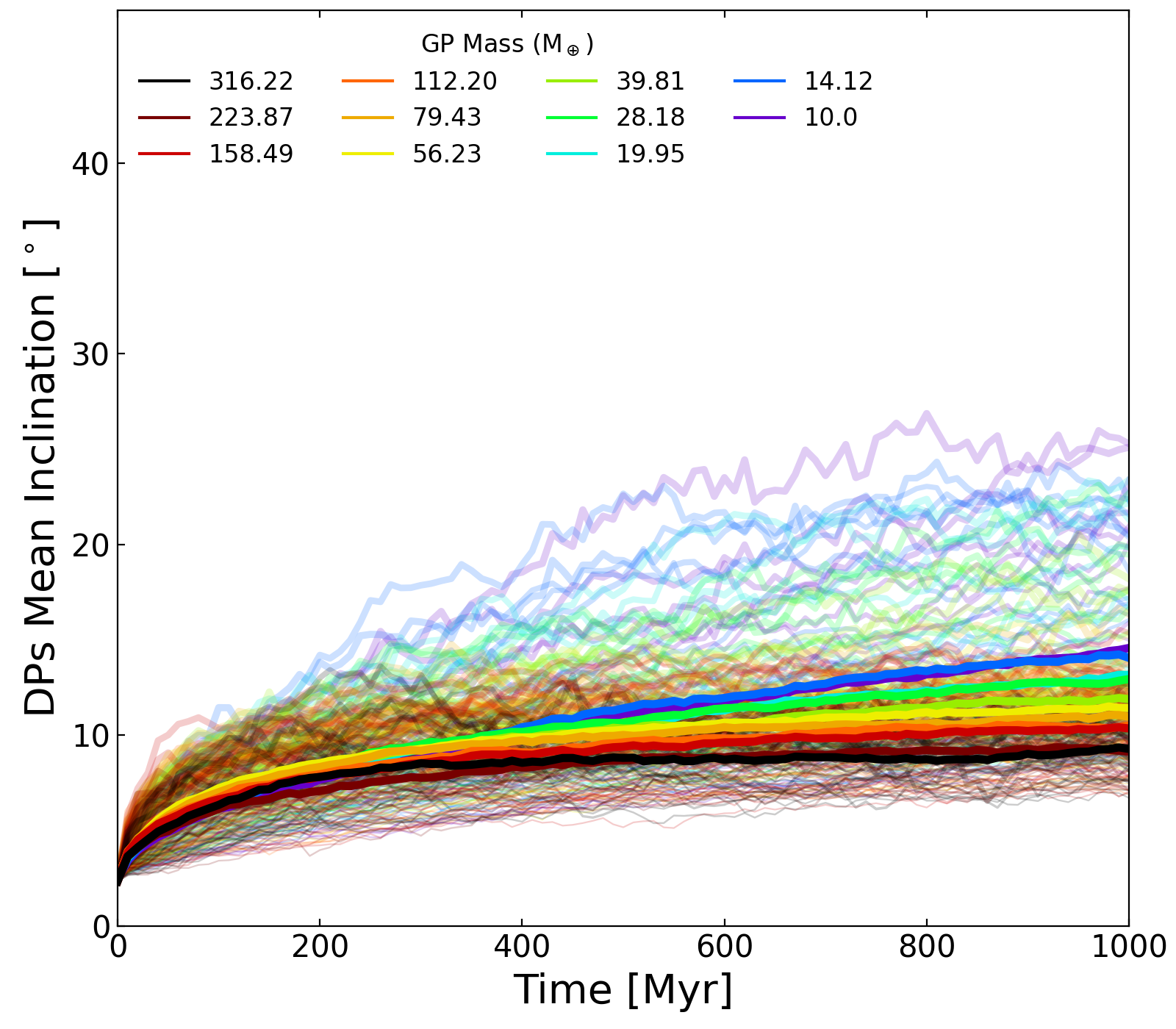} 
%        \caption{Inclination} 
    \end{subfigure}
    \caption{Same as \cref{fig:evol} but for the evolution of the survival fraction (top), mean eccentricity (middle), and inclination (bottom) of DPs in the discs. The color of the lines in both panels indicates the mass of the GP in the models, while line thickness represents the mass of the disc, with thicker lines corresponding to more massive discs (pale lines in the still frame, all but the last frame in the animated figure). The thickest lines in the still frame (and those of the last animated frame), correspond to the average of all disc masses for any given GP mass (An animated version of this figure can be found online at \href{https://figshare.com/projects/Mixed_Stirring_of_Debris_Discs/136118}{Figshare}).}
    \label{fig:DPsevol}
\end{figure}

We find that the evolution of massive DPs in the discs follows a similar trend to that of massless particles, but their self-stirring is slightly less efficient, as shown in \cref{fig:DPsevol} (cf. \cref{fig:evol}). There we present animations showing the evolution of the survival fraction (top panel), mean eccentricities (middle panel), and mean inclinations (bottom panel) for surviving DPs in the simulations as a function of time, in the same scheme as for the test particles in the previous sub-section. 

In the top panel of \cref{fig:DPsevol} we see the surviving fraction of DPs in each model system as a function of time. Again, consistent with the analogous plot for test particles in \cref{fig:evol}, we see that the DPs are more efficiently removed from the system with a more massive GP and a more massive disc. In the middle panel of \cref{fig:DPsevol} we can see trends in the behaviour of the DPs can be delineated for models with different GP masses, following the same general behaviour as for the test particles. The models with the lowest GP masses, below $15~\mathrm{M}_{\oplus}$, exhibit a rising mean eccentricity for the DPs up until the end point of our simulations. Models with GPs above that, but below $60~\mathrm{M}_{\oplus}$, again reach a plateau, and have a slow increase, but this time they have an obvious maximum before having a slow decrease in eccentricity at some point between 400~Myr and 1~Gyr. The time at which the highest value occurs, and its value are both dependent on the GP mass; more massive GPs have their maxima at earlier times and with lower mean eccentricity values. This is again a result of the increasing strength of interaction for DPs that more closely approach the GP. Furthermore, we see that overall the mean eccentricity of the DPs is lower than that of test particles. For models with GPs $> 60~\mathrm{M}_{\oplus}$ we again observe a third phase of evolution, in the first a rapid increase in mean eccentricity occurs, quickly reaching a peak within the first 200~Myr which is faster for more massive GPs; after this follows a decline, also faster the more massive the GP; finally, after the decline, a slow increment begins again until reaching an approximate steady state by the end of the simulations.

The maximum values of the average mean eccentricity for the models remain below $\simeq$0.35 for DPs (cf. 0.55 for test particles, which continue growing for the lowest mass GP models), with an apparent saturation limit at this value independently of GP mass. We can also see that the behaviour of the lines in \cref{fig:DPsevol} is noisier than in the case of test particles (\cref{fig:evol}), this is because the DP population is 10 times less numerous than the particles. We would expect this to also be true for any real disc since the number of DPs containing a substantial amount of the disc mass will always be a minority compared to the total population (starting with the largest bodies, which are the most dynamically relevant).

For any given GP mass there is a trend of larger eccentricities for larger disc masses. However, there is an overall dispersion for the evolution of each suite of simulations, and some individual simulations fall outside of the global trend e.g. the most massive discs for the systems with $28~\mathrm{M}_{\oplus}$ and $112~\mathrm{M}_{\oplus}$ GPs lie well above the other systems in their respective suites. These ``outliers'' may be attributed to stochastic events involving DP interactions or ejections influencing the overall evolution of that system.

The evolution of the mean inclination for DPs in our models is shown in the bottom panel of \cref{fig:DPsevol}. Again, we observe a similar behaviour to the one described above for the test particles, finding lower inclinations for larger GP masses, and also that the final inclination values are consistently lower. In this case, almost all the systems show a monotonic rise in mean inclination over the duration of the simulations with no turnover. Only the most massive GP systems ($> 220~\mathrm{M}_{\oplus}$) seem to reach a peak in their respective mean inclination within the duration of the simulations. Systems with lower mass GPs, $\mathrm{M}_{\rm GP} < 30~\mathrm{M}_{\oplus}$, are not yet slowing down at the end of the simulations. We also find that, for a given GP mass, more massive discs will produce larger mean inclinations. Overall, the greatest inclination values lie below 30$\degr$ for the DPs, regardless of GP mass, and take longer to undergo the same relative degree of excitation, as compared to the test particles in the same systems that can reach values close to 45$\degr$.

\subsection{Evolution of the Discs as a Whole}

\begin{figure*}
    \centering
    \begin{subfigure}[t]{0.33\textwidth}
        \centering
        \includegraphics[width=\linewidth]{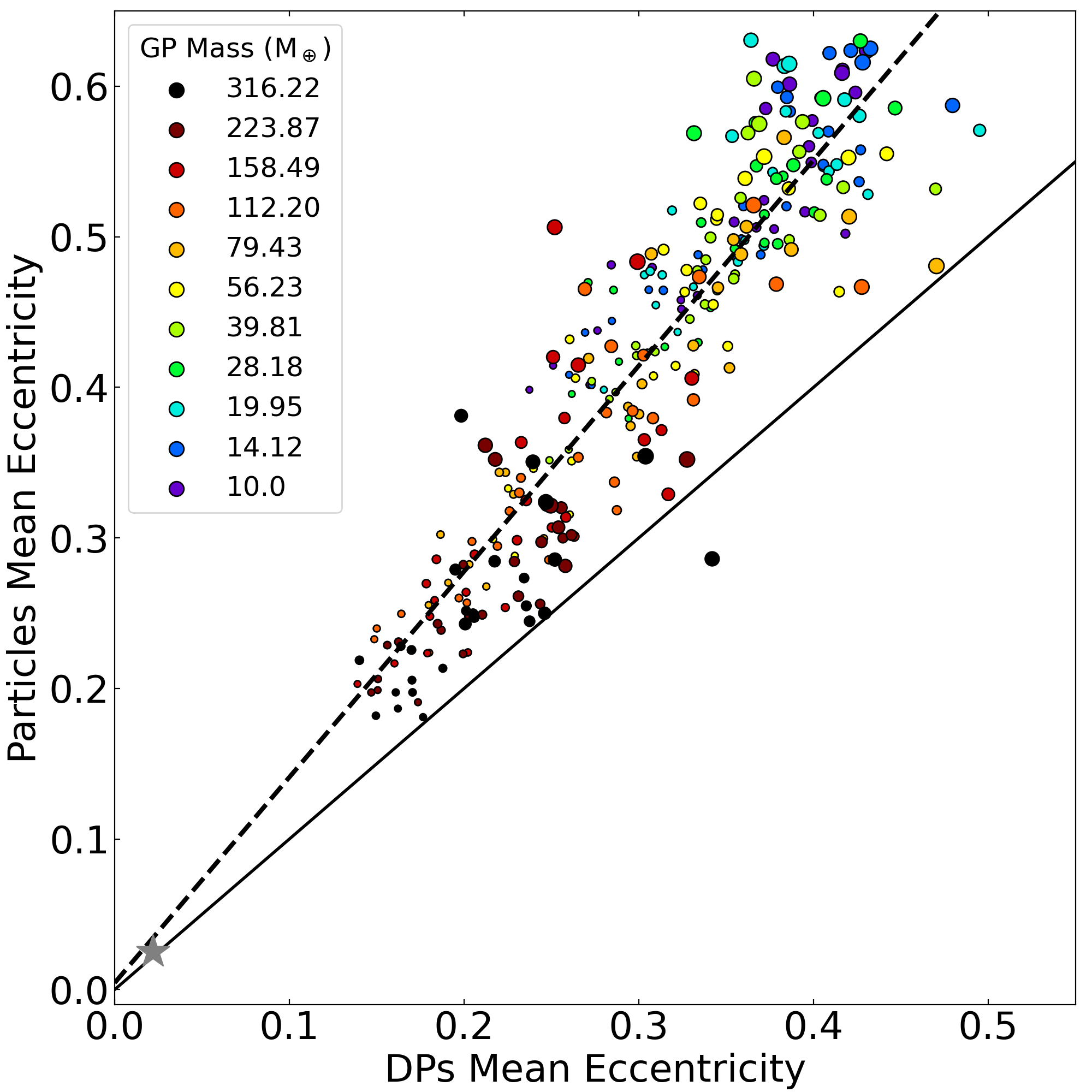} 
%        \caption{Eccentricity} 
    \end{subfigure}
    \begin{subfigure}[t]{0.33\textwidth}
        \centering
        \includegraphics[width=\linewidth]{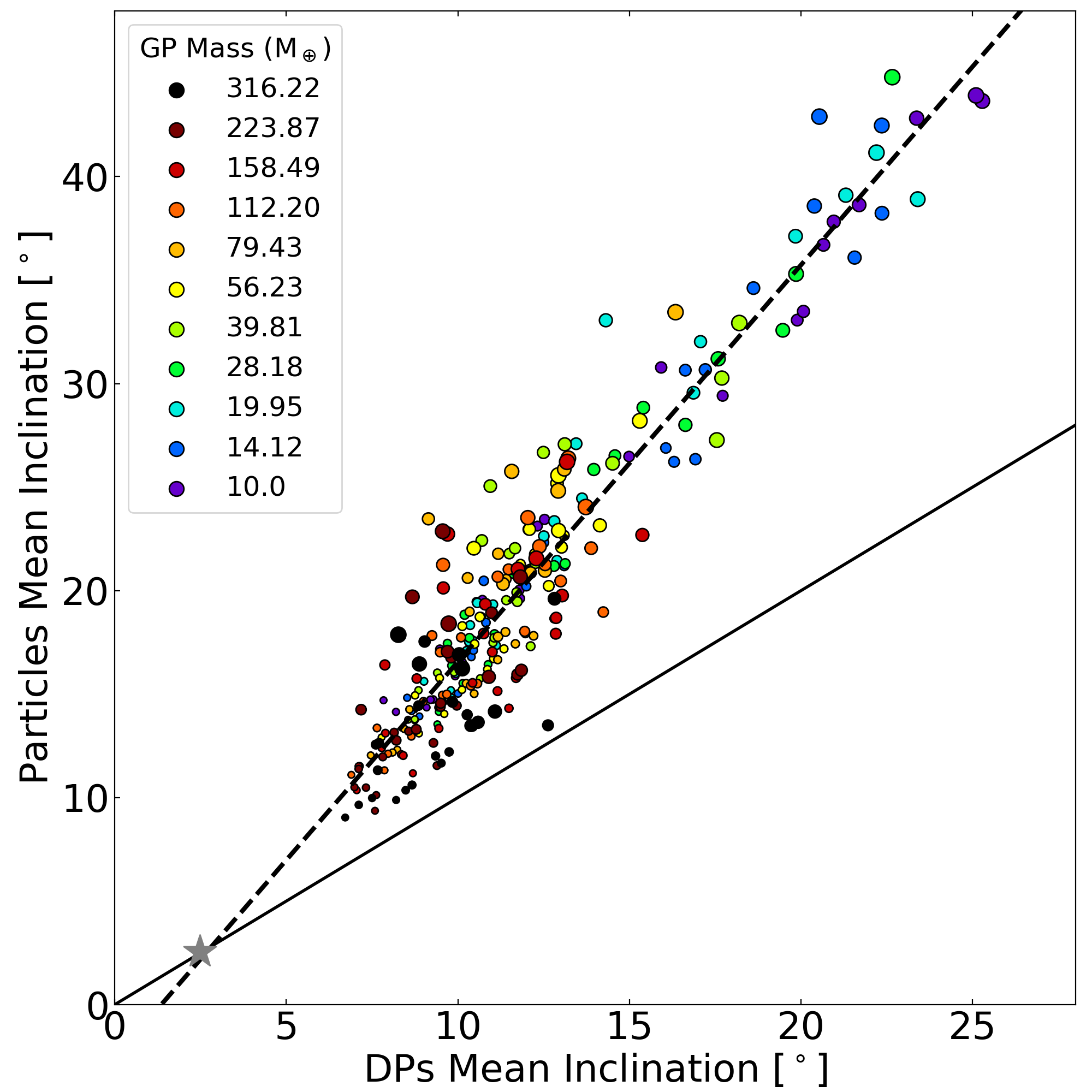} 
%        \caption{Inclination} 
    \end{subfigure}
        \begin{subfigure}[t]{0.33\textwidth}
        \centering
        \includegraphics[width=\linewidth]{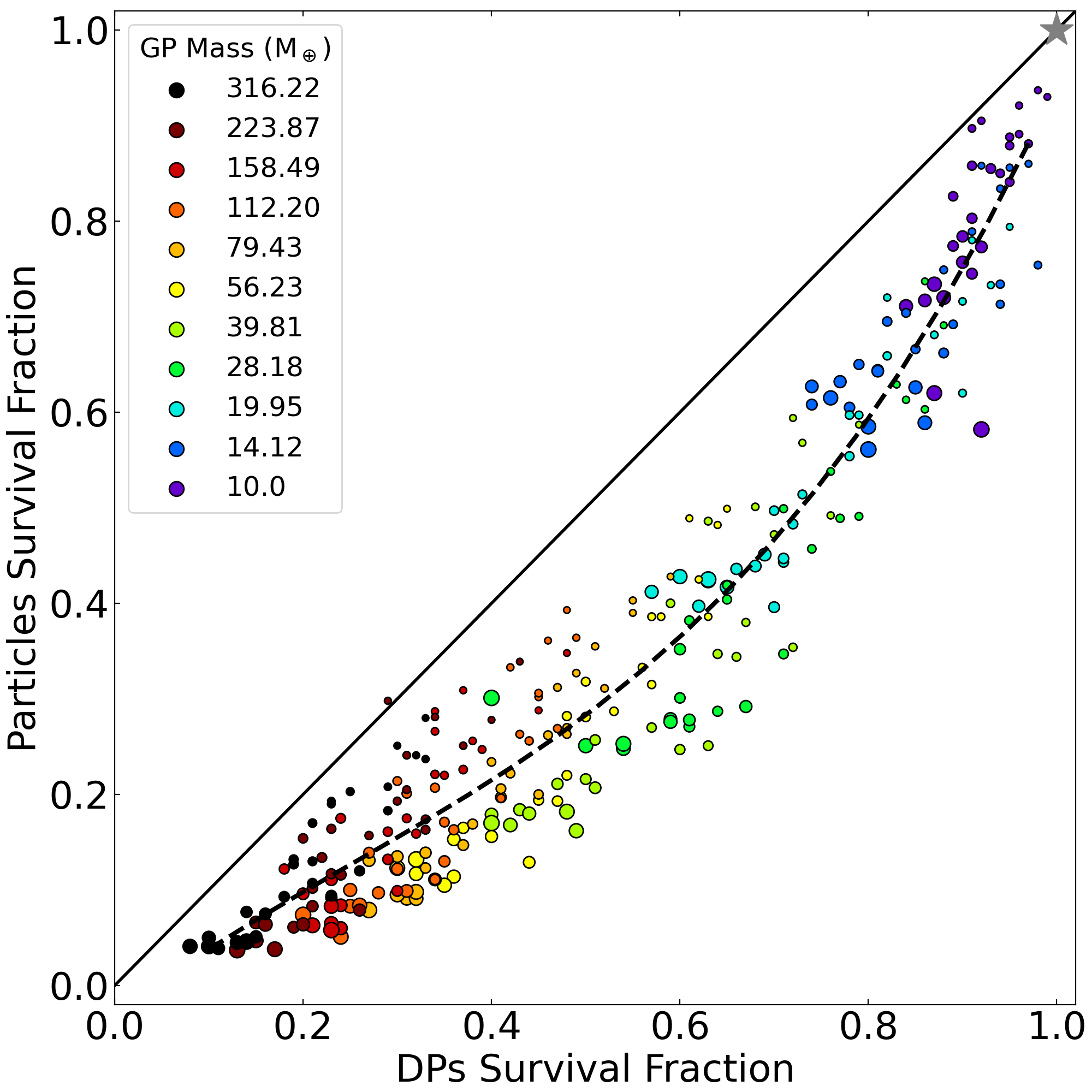} 
%        \caption{Inclination} 
    \end{subfigure}
    \caption{Comparison of the final mean orbital parameters and survival fractions of DPs {\it vs} test particles. The left panel shows the distribution of mean eccentricities, the middle panel for mean inclinations, and the right panel for survival fractions. The colors indicate the mass of the GP in the system, while the size of the dot represents the initial mass of the debris disc. The identity is indicated by the solid black line, while best fits are indicated by dashed black lines. Linear fits were done for both eccentricity and inclination, while a third-order polynomial was fitted to the survival fraction. The gray star in each panel represents the initial conditions of our systems.}
    \label{fig:DPsvsTPs}
\end{figure*}

To better understand the evolution of discs as complete systems, containing both massive and massless particles, as well as the relationship between the two, we begin by comparing the final values of the mean orbital parameters and survival fractions of test particles and DPs. In \cref{fig:DPsvsTPs} we show the final distribution of mean eccentricities (left panel), mean inclinations (middle panel), and survival fractions (right panel), of both populations, for all 286 systems; the different colors indicate the mass of the GP in that system, while the size of the dots represents the initial mass of the corresponding debris disc. 

In the left and middle panels of \cref{fig:DPsvsTPs} we see that for both mean eccentricity and inclination, the distribution of final values remains above the identity line (indicated by the solid black line) except for one outlier case in eccentricity which corresponds to one of the models with the most massive GP. We can see that the final conditions for all models closely follow a straight line. A comparison of the corresponding panels in \cref{fig:evol,fig:DPsevol}, shows that massless particles are more easily disturbed than DPs (as seen in \cref{fig:DPsvsTPs}); it can also be seen that the evolution of the eccentricity is much less mass dependent than that of the inclination. We applied a linear fit in both cases (dashed black lines in the left and middle panels of \cref{fig:DPsvsTPs}) to quantify how efficient the stirring of test particles is when compared to that of massive DPs. The best fit for the models in the eccentricity panel is given by $\left<e_{\it particles}\right>=1.366\left<e_{\it DPs}\right>+0.004$ and for the final inclination $\left<i_{\it particles}\right>=1.916\left<i_{\it DPs}\right>-2.602 \degr$; these fits show that the stirring of test particles is more efficient than that of DPs by factors of 1.366 for eccentricity and 1.916 for inclination. Both of these fits lie close to the gray star representing the initial conditions of all the distributions.

As in \cref{fig:evol,fig:DPsevol}, \cref{fig:DPsvsTPs} shows that the more massive discs (larger dots) are more efficient at stirring their particles than less massive ones (smaller dots) but that more massive GPs have a stabilising effect on the discs after a quick removal of the initially unstable minor bodies (both DPs and test particles); this comes about because massive GPs will tend to eject particles that pass close to them, whereas lighter GPs will perturb their orbits without ejecting them from the system.

The final distribution of survival fractions (right panel of \cref{fig:DPsvsTPs}) remains below the identity line, illustrating the greater difficulty for a planet in ejecting massive objects (DPs) than massless ones (test particles). A strong dependence on GP mass is observed in the final survival rate for both populations of minor bodies, demonstrating the efficiency of ejection. We find the relationship between the surviving fractions (${\it SF}_{\it particles}$ and ${\it SF}_{\it DPs}$) is best represented by a 3rd order polynomial of the form: 
\begin{eqnarray}
\label{eqn:trendSF}
{\it SF}_{\it particles}&=&0.932 \; {{\it SF}_{\it DPs}}^3 - 0.703 \; {{\it SF}_{\it DPs}}^2 \\
\nonumber
&&+ \; 0.746 \; {\it SF}_{\it DPs}-0.031.
\end{eqnarray}
As with eccentricity and inclination, extrapolation of this trend towards the less perturbed discs leads to the gray star representing the initial conditions; for survival fractions, there is also an obvious extrapolation to more violent systems and we find that our trend leads toward the (0,0) point where all particles would be ejected. The scatter of simulations around this trend line is generally more pronounced for the systems with higher GP masses (in the survival of both test particles and DPs). This is to be expected as it is interactions with the GP in each system that will dominate the removal of smaller bodies (either by collision or ejection). We find that the number of test particles removed by collisions remains approximately constant over the simulation grid, comprising about $2\%$ of the particles over the duration of each model run. By contrast, the number of ejection events is strongly correlated with the GP mass, with removals initially about $5\%$, and swiftly becoming greater by an order of magnitude or more with up to $95\%$ during a model run. As the GP mass decreases so too does the ejection efficiency, and they will only dynamically heat their companion discs rather than deplete them. This leads to a lower dispersion in the survival of DPs, but a comparable scatter in test particle ejection. The most massive GPs exhibit the tightest correlation with the observed trend. In these simulations, the GP rapidly stirs and depletes the disc (cf. \cref{fig:evol}) and if any minor body subsequently migrates into the perturbation region of the GP it is swiftly removed.

The most massive discs in the simulations for a given GP mass tend to lie below the trend line identified by \cref{eqn:trendSF}. This indicates segregation by disc mass within the distribution of surviving minor bodies, where the more massive (initial) discs are more depleted in both test particles and DPs for a given GP mass. This is the natural consequence of greater dynamical stirring by more massive individual DPs within the more massive disc for a given system, leading to particles (and DPs) passing into close interaction with the GPs. This tendency weakens and breaks down as the GP mass decreases, representing the decreasing capacity of the GP to deplete mass from the disc.

\begin{figure}
    \centering
    \includegraphics[width=\columnwidth]{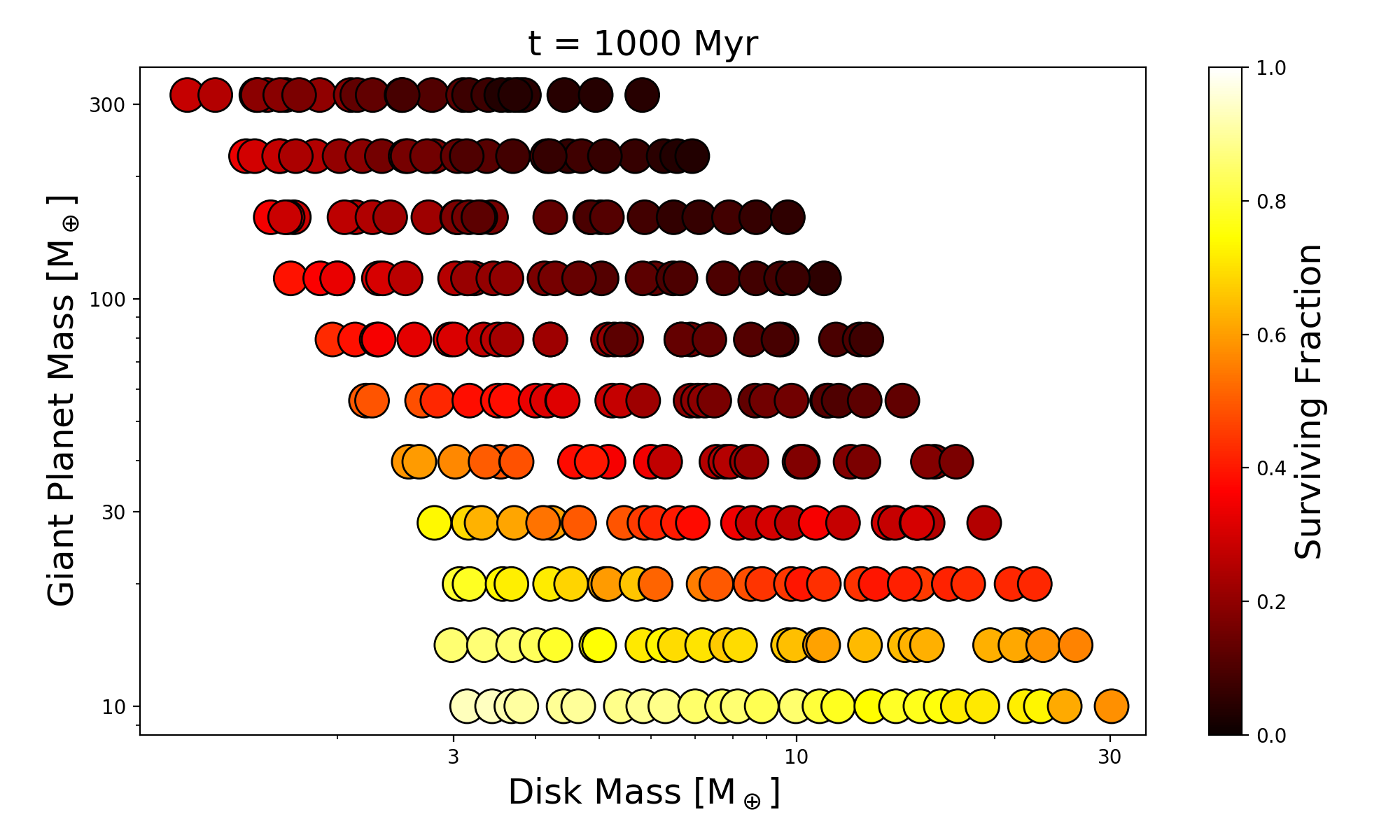}
    \includegraphics[width=\columnwidth]{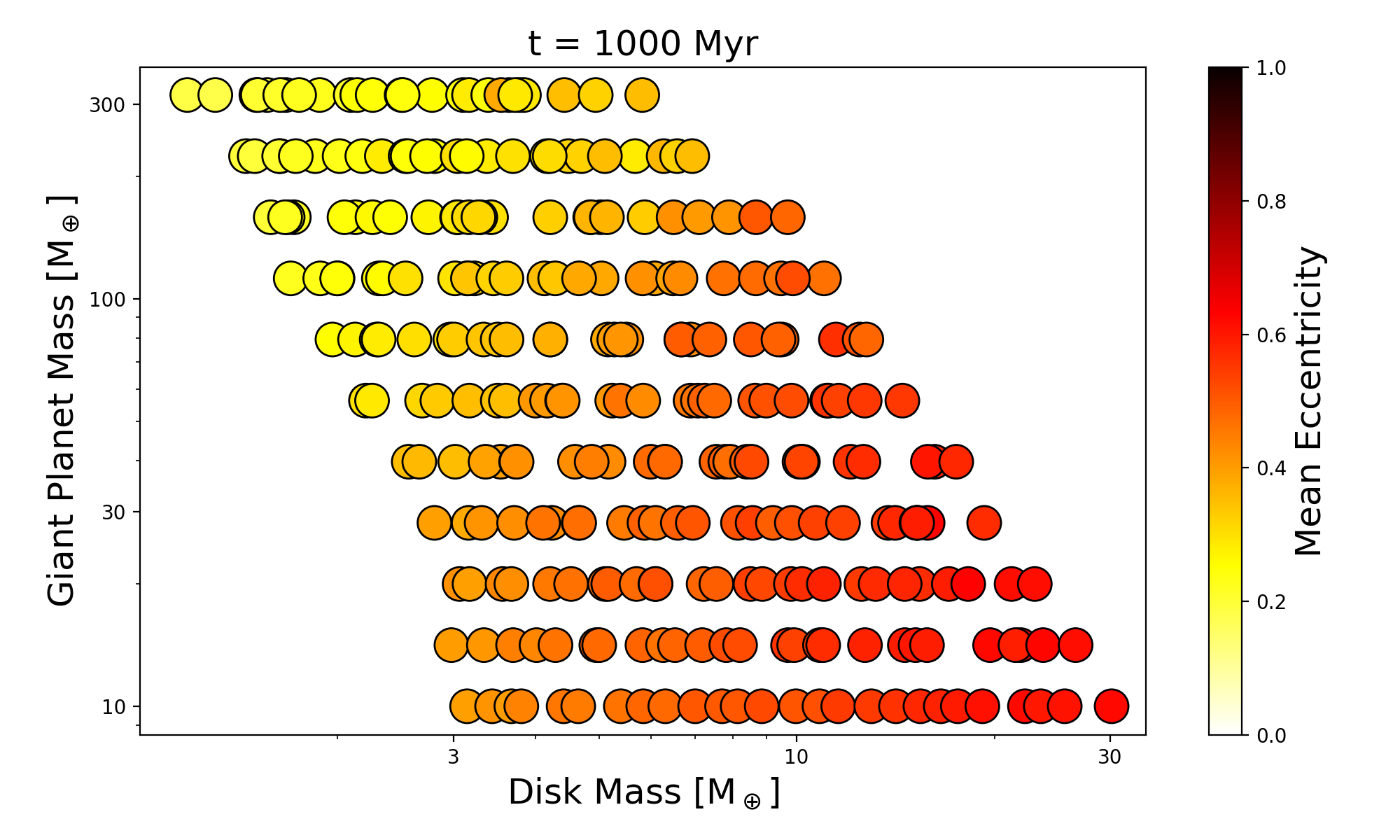}
    \includegraphics[width=\columnwidth]{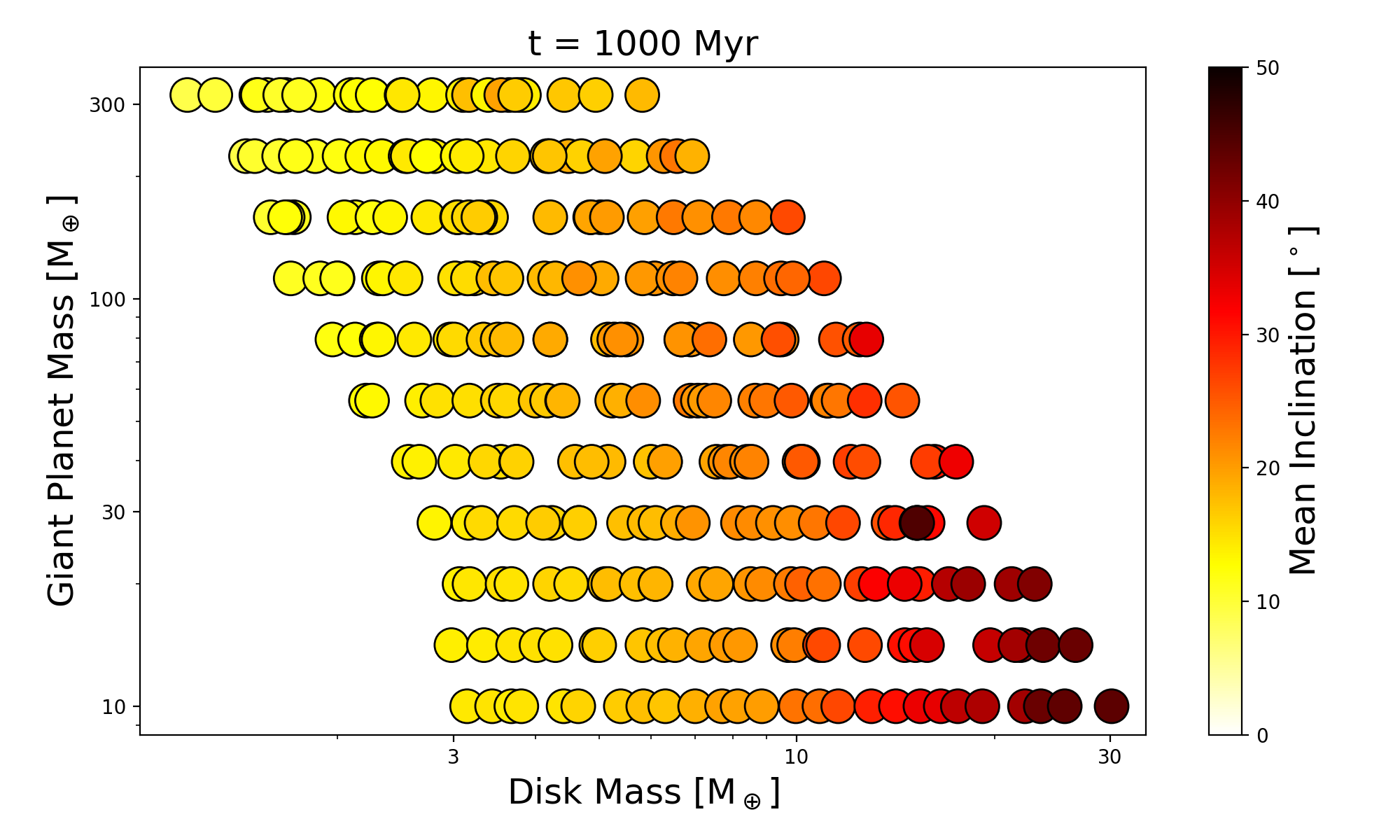}
    \caption{Animated figures illustrating the surviving fraction of test particles (top), mean eccentricity (middle) and mean inclination (bottom) as a function of the evolving disc mass vs. GP mass. The time step is in increments of 10~Myr (An animated version of this figure can be found online at \href{https://figshare.com/projects/Mixed_Stirring_of_Debris_Discs/136118}{Figshare}).}
    \label{fig:evolved_dd_mass}
\end{figure}

Most of the analysis of \cref{ssec:tpsevol,ssec:DPsevol} is focused on the point of view of the models, this is: we are classifying each model according to its initial conditions. However, this is not directly applicable to observations. From an observational point of view, it is more interesting to characterise a model according to its current parameters, and while the GP mass will not change, the disc mass will change with the ejection of DPs. Therefore, similar to the animated \cref{fig:evol}, in the animated version of \cref{fig:evolved_dd_mass} we show the time evolution of our 286 models by plotting the survival fractions, mean eccentricities, and mean inclinations of surviving particles at each time step, as a function of the evolving mass of the disc, instead of its initial mass.

In the top panel of \cref{fig:evolved_dd_mass} we show the survival fraction of test particles, where each snapshot corresponds to a 10~Myr evolution. The color of each circle indicates the particle survival fraction present in each disc at that time, with darker colours representing a lower survival fraction. 

We can see that the evolution of the survival rate is fastest for the most massive GP and (initial) disc mass combinations, with more than 50\% depletion of those discs occurring within the first few tens of Myr; in the same time frame, barely any ejections have occurred amongst the lower mass systems. By 100~Myr, systems with GP masses greater than $60~\mathrm{M}_{\oplus}$ have experienced substantial ejection, losing up to half the particles (but not necessarily half their mass), whereas systems below that have yet to experience any substantial ejections. At the 500~Myr point, the most massive systems have lost up to 90\% of their initial particles and only the least massive GP/disc systems are untouched by ejections. Beyond this time up to 1~Gyr the overall picture remains constant and the systems' evolution is more gradual.

If we focus on a fixed small area of the grid, instead of following the evolution of individual coloured circles, the behaviour of the survival fraction in that region becomes even more extreme, e.g., for a disc mass of $\sim 7~\mathrm{M}_{\oplus}$ the difference in survival fraction goes from $\approx 90\%$ (at low GP masses) to $\approx 10\%$ (at high GM masses) during the 1~Gyr simulation.

In the same sense, one should look at the animated middle panel of \cref{fig:evolved_dd_mass}, following the eccentricity evolution, as we looked at the animated top panel, i.e. we should focus on an area and not let our eyes drift away from it; by looking at a column centered at around $\sim 10~\mathrm{M}_{\oplus}$, we see that the eccentricity slowly rises with time; after the first few time steps the most eccentric models were those with a GP mass of $\sim 200~\mathrm{M}_{\oplus}$, but with time this maximum went all the way down to $10~\mathrm{M}_{\oplus}$ (although this took the best part of the 1~Gyr of our simulations). Another thing to note is that, while many individual dots reach saturation within our simulations, by looking at a fixed area we see that it keeps on evolving, mostly because simulations with more massive initial discs keep passing through our observation area (akin to the difference between Eulerian vs. Lagrangian evolution). By the end of our simulation, we find that there is a triangular region, in the lower right of the plot, that is mostly saturated with mean eccentricities $\sim 0.6$. This is quite extreme for discs that were initially dynamically cold with $\left<e_0\right>=0.025$, a $\sim$25 fold increase.

Finally, for the bottom panel of the animated \cref{fig:evolved_dd_mass}, following the inclination evolution, we observe that the evolution of inclination is slower than for eccentricity. After $\sim 100$~Myr the inclination is mostly homogeneous with only the most massive disc models showing signs of a significant stirring. During the next hundreds of Myr, a differentiation in the level of stirring becomes evident for individual columns, which seem to have uniform colors evolving in time, i.e., the excitation level for the inclination is more clearly dependent on the remaining debris disc mass than on the initial disc mass or the GP mass. At the end of the simulations, the maximum stirring has occurred for the most massive debris discs and the least massive GPs, however, since the ejection fraction grows with GP mass, as time passes, what would be an equally excited component in our most massive GP models has already been depleted. 

A skewed initial grid (rather than the rectangular one we considered here), with more massive debris discs for the more massive GP systems, might fill in some of this depleted parameter space. However, as the disc evolution timescale decreases with increasing disc mass, the observed regions of parameter space that are vacated in our simulations are necessarily void given the duration of the simulations. In this sense the structure we observe in our grid at 1~Gyr is not fixed; longer integration would necessarily drive all the systems to lower disc masses, leading to a more pronounced ``gap'' in the top right of these plots. This diagram, therefore, provides some constraints on the evolutionary pathway undertaken by observed debris discs with the constraints of the stellar age and inferred disc mass.

\subsection{Evolution of GPs}

\begin{figure*}
    \centering
    \begin{subfigure}[t]{0.33\textwidth}
        \centering
        \includegraphics[width=\linewidth]{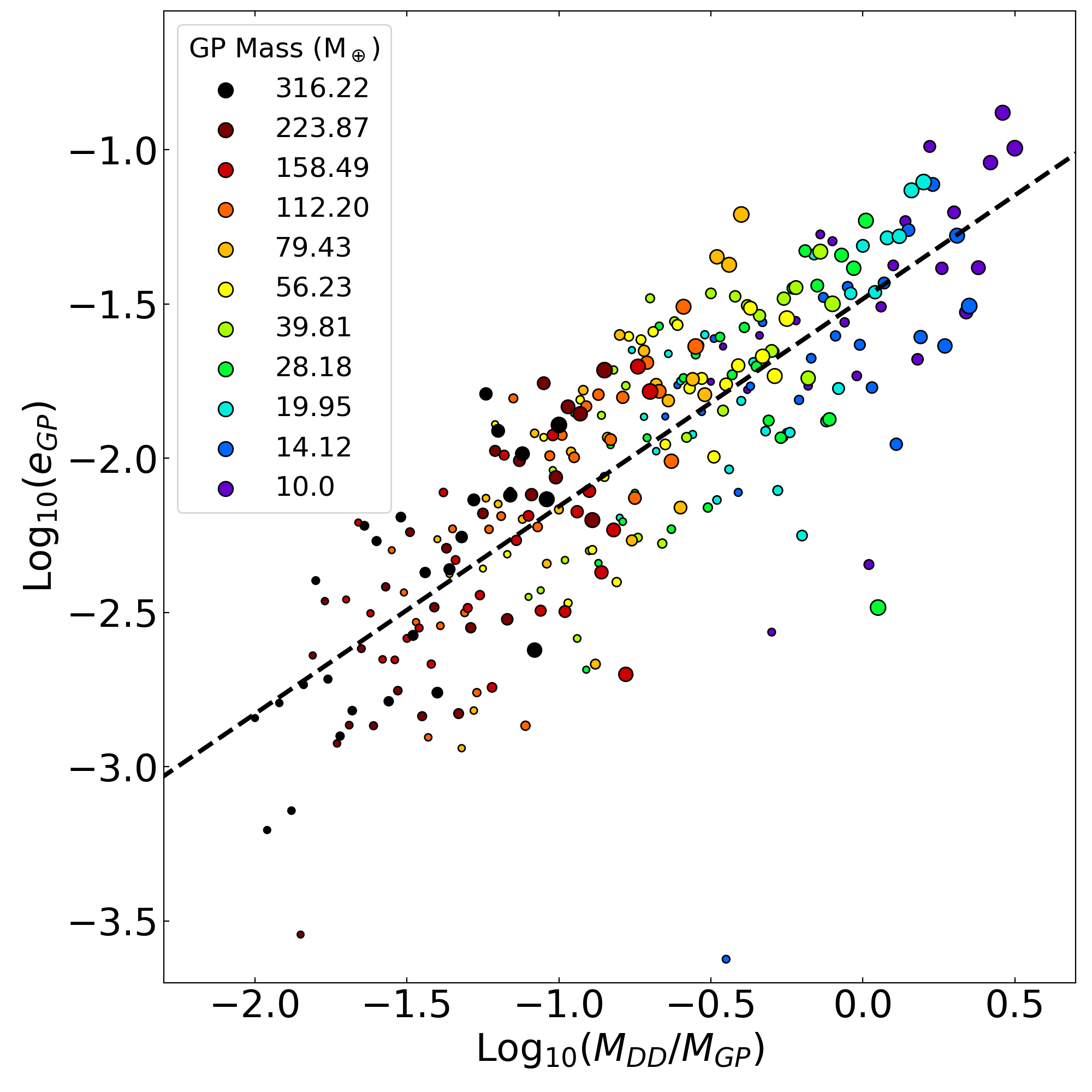} 
%        \caption{Eccentricity} \label{sfig:e}
    \end{subfigure}
    \begin{subfigure}[t]{0.33\textwidth}
        \centering
        \includegraphics[width=\linewidth]{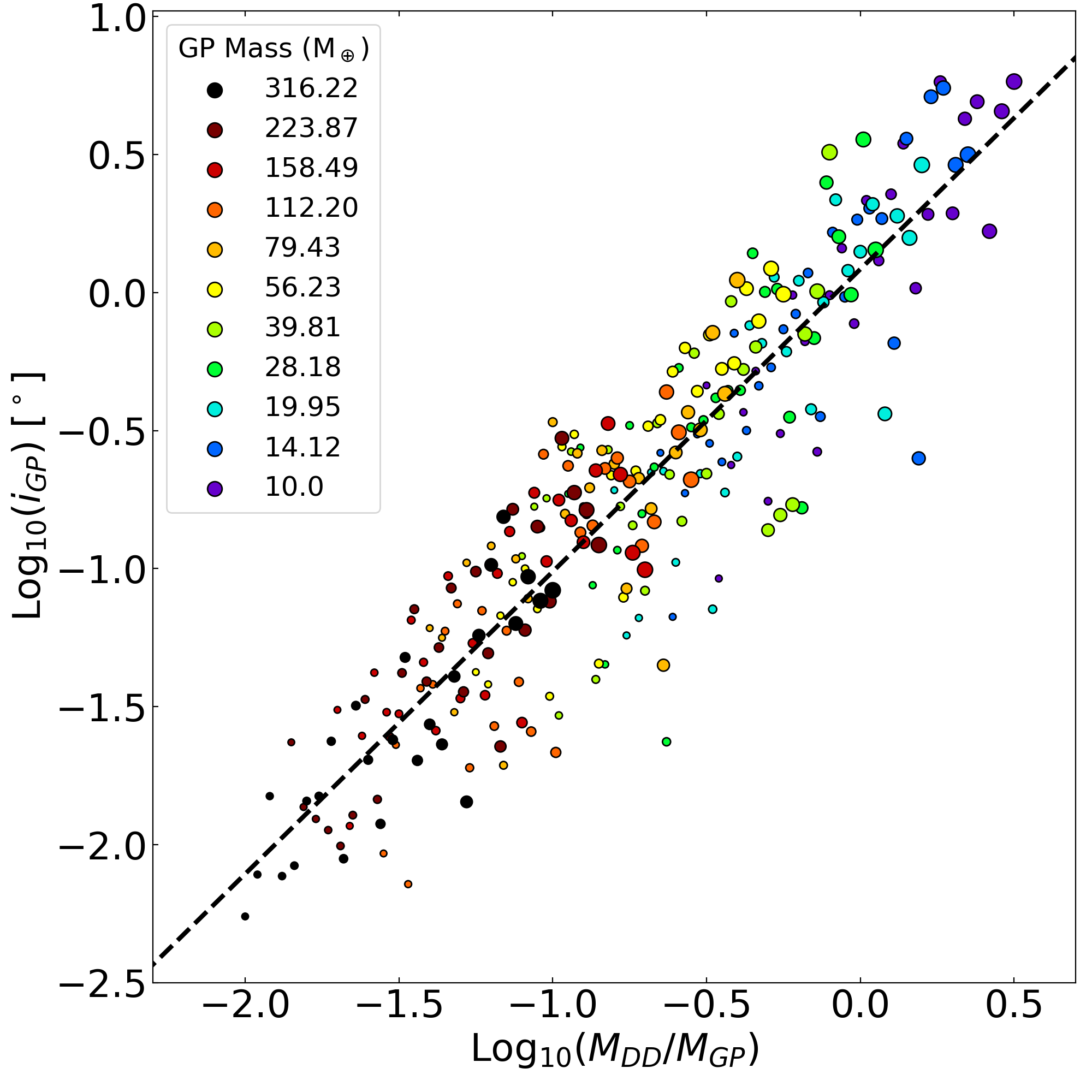} 
%        \caption{Inclination} \label{sfig:inc}
    \end{subfigure}
        \begin{subfigure}[t]{0.33\textwidth}
        \centering
        \includegraphics[width=\linewidth]{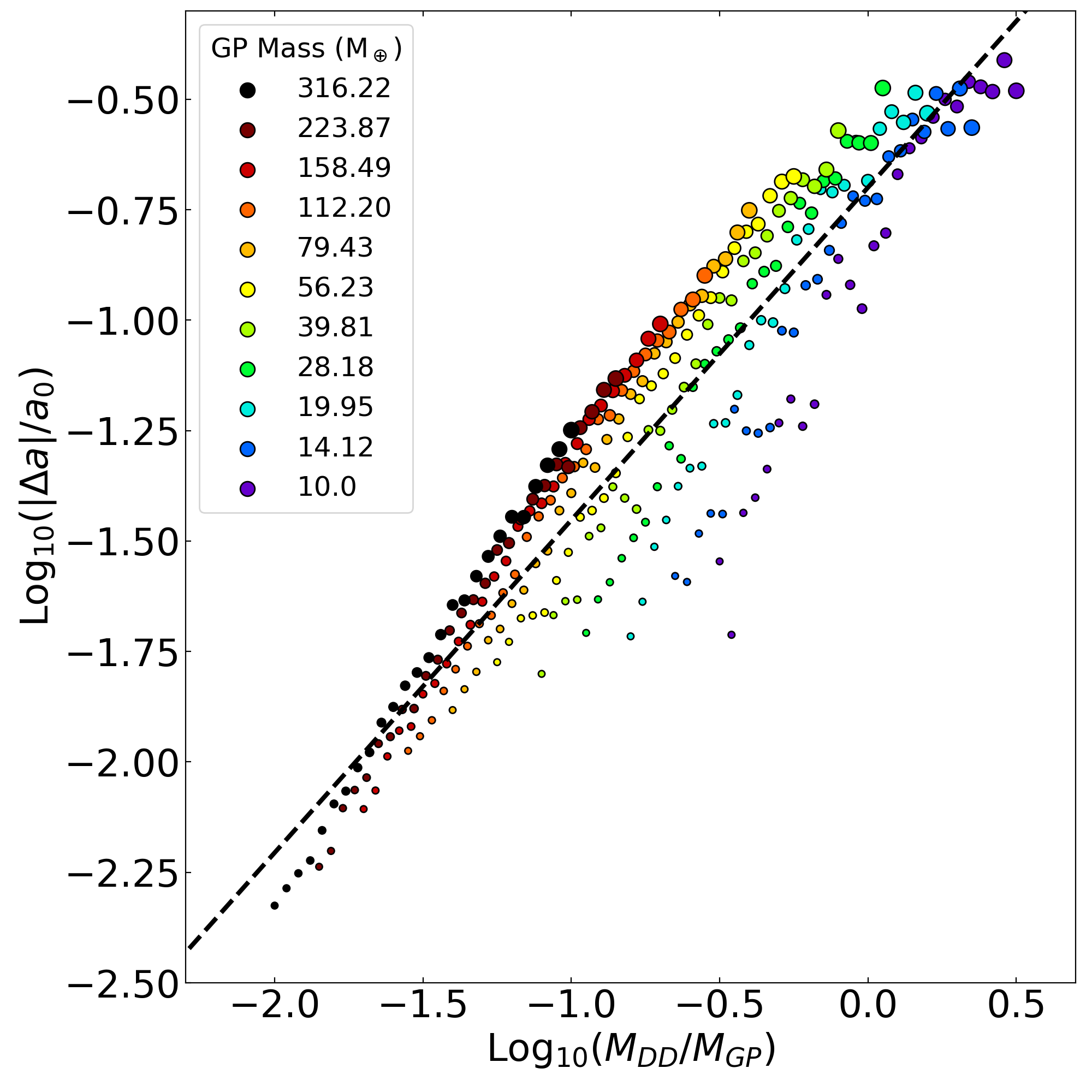} 
%        \caption{Inclination} \label{sfig:inc}
    \end{subfigure}
    \caption{Final orbital values for the GPs as a function of the mass ratio. The left panel shows the distribution of final eccentricities, the middle panel for final inclinations, and the right panel for final semimajor axes changes. As in \cref{fig:DPsvsTPs}, the colors indicate the mass of the GP in the system, while the size of the dot indicates the initial mass of the debris disc. Linear fits in these Log-Log planes are indicated by the dashed lines (see text for details).}
    \label{fig:GPsEvol}
\end{figure*}

Besides the evolution of the debris disc systems as a whole, the GPs in our models experience  modifications to their initial orbital parameters; this is due to the interactions between the GP and massive DPs, which results in the interchange of angular momentum that leads to an overall increase in their eccentricity and ultimately to ejections of some DPs. Although small in most cases, the orbital perturbations experienced by some of the GPs in our models can be significant; specifically: large inward orbital drifts, of up to 10~au, are observed in systems with the less massive GPs and the most massive debris discs, i.e. in those systems with the largest mass ratio, as given by $M_{\it DD}/M_{\it GP}$.

In \cref{fig:GPsEvol} we only show the final distribution, in logarithmic values, for the eccentricities (left panel), inclinations (middle panel), and semimajor axis changes (right panel) of the GPs in our 286 models, as a function of the logarithm of the mass ratio of the system. In log-log space, those three distributions can be well described by linear fits. 

At the end of the simulations we found that most of our GPs would be considered to have remained in cold orbits (only 3, out of 286, have $e>0.1$, while only 5 have $i>5^\circ$); however, about 30\% of the GPs in our simulations have lost a significant fraction of their angular momentum, having a noticeable decrease in their semimajor axis by the 1 Gyr mark, $a<0.9a_0$.

At any point during the simulations, the three distributions ($e$, $i$, and $\left|\Delta a\right|/a_0$) can be well described by linear fits that slowly evolve with time, with both the absolute value as well as the mass fraction dependence slowly increasing. By fitting all simulations at 100 Myr intervals, and subsequently fitting a time dependence to the linear fits we obtain:
\begin{align}
\nonumber {\log_{10}}(e_{\it GP}) = \; & 0.3383 \left({\frac{T}{\rm Myr}}\right)^{0.1005} \,{\log_{10}}(M_{\it DD}/M_{\it GP}) \\
& + 0.2061\,{\log_{10}}\left({\frac{T}{\rm Myr}}\right)-2.0871, \\
\nonumber {\log_{10}}(i_{\it GP}) = \; & 0.6454 \left({\frac{T}{\rm Myr}}\right)^{0.0751} \,{\log_{10}}(M_{\it DD}/M_{\it GP})\\
& + 0.3265\,{\log_{10}}\left({\frac{T}{\rm Myr}}\right)-0.9025, \\
\nonumber {\log_{10}}(\left|\Delta a\right|/a{_0}) = \; & 0.3138 \left({\frac{T}{\rm Myr}}\right)^{0.1281} \,{\log_{10}}(M_{\it DD}/M_{\it GP})\\
& + 0.4582\,{\log_{10}}\left({\frac{T}{\rm Myr}}\right)-2.0646, 
\end{align}
for eccentricity, inclination, and semimajor axis change, respectively. 

\section{Discussion}
\label{sec:discussion}

In this work, we did not focus on the sculpting process of the edges of the discs, nor on the disc shapes; this is why we choose 10 Hill radii as the inner edge of the discs and not 5 Hill radii as been done elsewhere \citep[e.g.][]{Pearce14}. We further assumed a GP in a circular and planar orbit to minimise the impact of the planet on the disc. We instead focused on the stirring process produced as a result of the interaction of the massive DPs embedded in the debris discs, but such a process was somewhat dominated by the presence of the GP. We focused on determining the stirring levels as functions of both the GP and debris disc masses (assuming the shapes and disc edges are imprinted on the debris discs by the giant planetary companion).

Our model spans GP masses between 10 $\mathrm{M}_{\oplus}$ (approximately 60\% the mass of Neptune) and 316 $\mathrm{M}_{\oplus}$ (approximately the mass of Jupiter); \cite{Pearce22} estimate that Neptune to Saturn-mass planets are the minimum needed to stir most of their 178 modeled discs (though some needing Jupiter mass planets, assuming maximum eccentricities of 0.3). Similarly, the range of disc masses in this analysis, 3.16 to 31.6 $\mathrm{M}_{\oplus}$, are consistent with expectations based on both observations and theoretical considerations \citep{Mulders21,KrivovWyatt21}. Several other studies predict larger masses ($> 100~\mathrm{M}_{\oplus}$) in order for debris discs to be self-stirred \citep[e.g.][]{Krivov18,KrivovWyatt21}. Nonetheless, in this work, we found that small masses in debris discs can result in large stirring values, up to a 25-fold increase in the most extreme cases. Thus an efficient stirring is possible for small disc masses ($< 10~\mathrm{M}_{\oplus}$), if ever perhaps containing larger than expected perturbers, as some of the DPs present in the most massive discs we considered have assigned masses close to 1~$\mathrm{M}_{\oplus}$.

Our massive objects are initially thought to be real `dwarf planets' (DPs), as long as we adopt the definition of DP as an object that has not cleared its neighborhood from debris (yet). We could expect massive debris discs (much more massive than our Kuiper belt) to contain more massive objects, though this is not necessarily true, depending on planetary formation mechanisms, disc mass density, etc. Recent studies on dust formation and excitation place limits to the most massive objects present in massive debris discs to be around 5 times the mass of Pluto. Based on spatially resolved observations of the vertical scale heights of the debris discs around AU Mic and $\beta$ Pic, the most massive bodies present in those discs could be up to 9$\times10^{-5}$ and 0.4 M$_\oplus$, respectively \citep{2019Daley,2019Matra}. However, more massive objects might be present in debris discs (without leaving a piece of observational evidence, such as bumps or gaps), if we assume the mass range in planetesimals scales linearly with the overall mass of the disc.

Limiting the mass of the DPs to be similar to the mass of Pluto would diminish the stirring effect in both $e$ and $i$. According to shorter duration simulations ($t = 50~$Myr) with 41, 100, and 250 DPs (and 410, 1\,000, and 2\,500 test particles) we ran as consistency checks, this correction should be approximately a factor of 1.5, and definitely less than a factor of 2. 

Notably, by using {\sc Mercury}, the problem becomes computationally intractable when considering more than a few hundred massive DPs; new tools are required to expand the grid with a greater number of DPs and particles, such as GPU-based simulations. We leave this question open for future work.

The main stage for excitation evolution in our simulations occurs in timescales of the order of a few and up to 100 Myr; while the time required for our systems to acquire their final configurations, i.e. reach their saturation levels, is of the order of 150~Myr to more than 1~Gyr scales. These timescales are similar to the ages of host stars for many observed debris disc systems; thus we would expect that many of the observed systems with similar physical parameters as those covered in this work, would already be settled in their final configuration, currently experiencing a quiet steady-state evolution.

The timescales derived here are a function of the chosen architecture of the model, adopting a 1~$\mathrm{M}_{\odot}$ star with a planetary companion at 30~au and a Kuiper belt-like disc beyond that. However, the evolutionary timescales can be easily scaled for different stellar masses or disc semi-major axes, as the dynamics should be self-similar, provided physical collisions are a negligible cause of removal of bodies. For a different central star mass, all the masses should scale proportional to the new stellar mass, and the timescales should be modified as the inverse of the square root of the mass; for a different GP orbital radius, the masses should not be modified, and the timescales should be modified as the mass to the -3/2 power. For the Vega system, with a stellar mass $\sim$2~$\mathrm{M}_{\odot}$ and a planetesimal belt around 100~au \citep{Matra20,Marshall22}, the equivalent timescale would be nearly three times longer than the evolution of the models considered here (depending on the exact location of the GP used in Vega). 

A debris disc is the result of a collisional cascade within a planetesimal belt triggered by dynamical excitation, either intrinsically by the largest planetesimals within the belt \citep{Krivov18} or extrinsically by an external perturber \citep[e.g.][]{Mustill09}. The range of relative velocities among planetesimals, required to trigger the onset of the collisional cascade, is typically estimated as 100 to 300 m/s \citep[e.g.][]{KenyonBromley01}. On the other hand, such relative velocities can be estimated from the average orbital parameters of the dust-producing small objects in the discs, as $V_{\it rel}=V_K\sqrt{1.25e^2+I^2}$, where $V_K$ is the Keplerian velocity at the distance $a$ from the star \citep{Lissauer93,WyattDent02}. \citet{Krivov18} argued though, that the average inclinations are not terribly important when determining the relative velocities among planetesimals, since eccentricities grow much faster than inclinations in debris disc models. Thus, one can simply estimate the relative velocities from the root mean square eccentricity of the planetesimals as $V_{\it rel}=V_K\sqrt{\left<e^2\right>}$. In any case, an estimation of such relative velocities in all of our models shows that values close to 1 km/s are quickly reached, in less than 10 Myr, regardless of the initial debris disc mass or the GP mass of the model. Indeed, velocities of collisions within the belt modeled here, range from $\sim$250 m/s to $\sim$2000 m/s, therefore locating themselves safely on the side of a collisional cascade capable of producing dust. 

A population of planetesimals on eccentric orbits within a debris disc would produce a halo of millimetre dust grains. Such structures have been identified in ALMA observations of several systems, including HR~8799 \citep{Geiler19}, HD~32997 and HD~61005 \citep{2018Macgregor}. The typical eccentricity of dust grains within debris discs inferred from their spatially resolved belts lies in the range 0.1 to 0.3 \citep[based on 11 discs, see figure 9 of][]{2021Marino}; this level of eccentricity is consistent with the mean eccentricity induced by the dwarf planets in this set of simulations.

\section{Summary and Conclusions}
\label{sec:conclusions}

In this work, we performed a suite of 286 numerical simulations to explore the stirring effects that a combination of giant and dwarf planetary perturbations would have on the long-term evolution of initially cold debris disc models. Our systems are formed by a solar mass star, a giant planet initially located at 30 au in a circular and planar orbit, and 100 massive dwarf planets embedded in a disc described by 1000 test particles. The orbital distribution of the discs was drawn randomly for small values of eccentricity (between 0 and 0.05) and inclination (between 0$^\circ$ and 5$^\circ$). We initially located the inner edge of our discs at 10 Hill radii from the GP, with a total width of 30 au. Our 1 Gyr long simulations take into account the perturbations from the GP and the DPs over test particles and among themselves. 

The evolution timescale for the eccentricity and inclination depends mostly on GP mass, where the simulations with more massive GPs evolved faster than those with less massive GPs. On the other hand, the limit to the heating depends on both GP mass and disc mass, with large disc masses and small GP masses being able to heat the disc more than simulations with light discs and/or heavy GPs. Part of the reason why massive GPs are less efficient at heating the disc is their tendency of ejecting ``warm'' particles before they can get extreme values of either eccentricity or inclination.

In all models the mean inclination rises quickly (or at least relatively quickly) before slowing down, only the most massive GPs seem to be able to level off before the 1~Gyr mark. The eccentricity evolves faster with many of the simulations reaching a plateau before the end of the simulation. Very massive GPs heat their discs very quickly which then slowly cool down by ejecting the more excited particles. The effect on the eccentricity is larger than for the inclination.

Massless particles, which in real systems could be considered as the less massive members of the discs, such as cometary nucleii, are more mobile than massive objects (DPs), therefore they become 'hotter', i.e. more eccentric, more inclined, and are easier to be ejected (they have a poorer survival rate). Nonetheless, DPs reach significant stirring levels as well and have only slightly better chances of survival than test particles.

The values of both eccentricity and inclination for test particles at a given time have a better correlation with the remaining mass of the debris discs than with the GP mass or the initial debris disc masses; this is particularly evident for the inclination.

GPs themselves are perturbed by their interactions with massive DPs, the most significant perturbations occur when the mass of the disc is comparable to the mass of the GP. In such cases, a significant  inward migration of the GP takes place, of up to $\sim10$ au, leaving a stirred disc that is not able to cool off by ejecting ``warm'' particles, with a far away GP closer to its star. 

The masses in debris discs explored in this work, and specifically their evolving remaining masses, are indeed very small when compared to those expected to be able to stir the disc by the self-stirring scenarios \citep{KrivovWyatt21,Krivov18}, but here we highlight the fact that even with such small masses, which involve a small number of massive perturbers (100 DPs initially), and perhaps more importantly, not-so-massive objects, are capable of increasing in an important percent, while acting together with the GP, the eccentricities and inclinations of debris disc particles. This result is similar to the enhancement of cometary production in the Kuiper belt found by \citet{Munoz19} and could have additional implications for the production of exocomets in extrasolar planetary systems. 

Taking everything into account, we have found that a combination of perturbers, consisting of embedded dwarf and external giant planetary masses, is in general more efficient in the stirring of cold debris discs than one or the other mechanism acting independently. 

\section*{Data Availability}

The data underlying this article are available in the article and in its online supplementary material. The animations, supplementary data, and analysis scripts are provided for public access on  \href{https://figshare.com/projects/Mixed_Stirring_of_Debris_Discs/136118}{Figshare}.

\section*{Acknowledgements}

The authors thank the referee, Alex Mustill, for his constructive and helpful comments. JPM acknowledges research support by the Ministry of Science and Technology of Taiwan under grants MOST107-2119-M-001-031-MY3 and MOST109-2112-M-001-036-MY3, and Academia Sinica under grant AS-IA-106-M03.

\textit{Software:} This work has made use of the symplectic integrator package {\sc mercury} \citep{Chambers99}, and the {\sc Python} modules {\sc Matplotlib} \citep{Hunter07}, and {\sc NumPy} \citep{Harris20}.



\bibliographystyle{mnras}
\bibliography{ddsbib} % if your bibtex file is called example.bib

%%%%%%%%%%%%%%%%% APPENDICES %%%%%%%%%%%%%%%%%%%%%

%\appendix


%If you want to present additional material

%%%%%%%%%%%%%%%%%%%%%%%%%%%%%%%%%%%%%%%%%%%%%%%%%%

% Don't change these lines
\bsp	% typesetting comment
\label{lastpage}
\end{document}